\documentclass{lmcs}
\pdfoutput=1

\usepackage{lastpage}
\lmcsdoi{22}{4}{15}
\lmcsheading{}{\pageref{LastPage}}{}{}%
{Jul.~15,~2025}{Nov.~05,~2026}{}

\usepackage[utf8]{inputenc}
\usepackage[T1]{fontenc}
\usepackage[pdfencoding=auto, psdextra]{hyperref}
\gundef\bibsetup
\usepackage[style=alphabetic,natbib=true]{biblatex}
\bibliography{mybib}


\usepackage{microtype}
\usepackage{bm}
\usepackage{amsmath,amssymb}
\usepackage{booktabs,array,csquotes}
\usepackage{mathtools, url, stmaryrd, etoolbox}
\usepackage[english]{babel}
\usepackage{xcolor}

\usepackage{mathpartir}

\usepackage{listings}

\newcommand{\ang}[1]{{\left<#1\right>}}
\newcommand{\eqv}[1][]{{\hspace{2pt}=_{#1}\hspace{2pt}}}
\newcommand{\snocop}{\mathbin{\mkern0.5mu\mathord{:}\mkern-1.5mu{\raisebox{0.1mm}{\scriptsize <}}\mkern-0.5mu}}

\newcommand {\FV}{\textsf{FV}}
\newcommand {\Vars}{\textsf{Vars}}
\newcommand {\app}{\raisebox{0.3 ex}{\scalebox{0.8}{ $\bm{++}$ }}}
\newcommand {\subst}{{\texttt{subst}}}
\newcommand {\simplesubst}{{\texttt{ssubst}}}
\newcommand {\vcrefresh}{{\texttt{vcRefresh}}}

\newcommand\code[1]{{\tt\small #1}}
\definecolor{dkgreen}{rgb}{0,0.3,0.3}
\definecolor{ltblue}{rgb}{0,0.3,0.3}
\definecolor{dkviolet}{rgb}{0.3,0,0.5}
\definecolor{dkred}{rgb}{0.4,0,0.0}
\definecolor{dkblue}{rgb}{0.0,0,0.0}

\newcommand{\rocqsource}[3]{%
  \href{https://github.com/kalmera/lambda-simple-alpha/blob/aug26/#1\#L#2}{%
    \raisebox{-0.08em}{\includegraphics[height=0.72em]{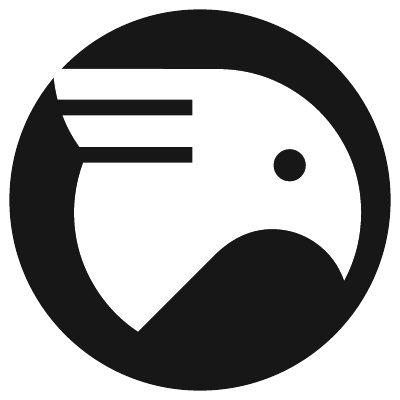}}%
    \nobreakspace{#3}}}

\newcommand{\rocqrepo}[1]{%
  \href{https://github.com/kalmera/lambda-simple-alpha/blob/aug26/}{%
    \raisebox{-0.08em}{\includegraphics[height=0.72em]{graphics/rocq-github-mark.pdf}}%
    \nobreakspace{#1}}}

\lstdefinelanguage{Coq}{
    mathescape=true,
    texcl=false,
    morekeywords=[1]{Section, Module, End, Require, Import, Export,
        Variable, Variables, Parameter, Parameters, Axiom, Hypothesis,
        Hypotheses, Notation, Local, Tactic, Reserved, Scope, Open, Close,
        Bind, Delimit, Definition, Let, Ltac, Fixpoint, CoFixpoint, Add,
        Morphism, Relation, Implicit, Arguments, Unset, Contextual,
        Strict, Prenex, Implicits, Inductive, CoInductive, Record,
        Structure, Canonical, Coercion, Context, Class, Global, Instance,
        Program, Infix, Theorem, Lemma, Corollary, Proposition, Fact,
        Remark, Example, Proof, Goal, Save, Qed, Defined, Hint, Resolve,
        Rewrite, View, Search, Show, Print, Printing, All, Eval, Check,
        Projections, inside, outside, Def, where, Property, Equations, Equations?},
    morekeywords=[2]{forall, exists, exists2, fun, fix, cofix, struct,
        match, with, end, as, in, return, let, if, is, then, else, for, of,
        nosimpl, when, into},
    morekeywords=[3]{Type, Prop, Set, true, false, option},
    morekeywords=[4]{pose, set, move, case, elim, apply, eapply, clear, hnf,
        intro, intros, generalize, rename, pattern, after, destruct,
        induction, using, refine, inversion, injection, rewrite, congr,
        unlock, compute, ring, field, fourier, replace, fold, unfold,
        change, cutrewrite, simpl, have, suff, wlog, suffices, without,
        loss, nat_norm, assert, cut, trivial, revert, bool_congr, nat_congr,
        symmetry, transitivity, auto, split, left, right, autorewrite,
        dependent, constructor, simp},
    morekeywords=[5]{by, done, exact, reflexivity, tauto, romega, omega,
        assumption, solve, contradiction, discriminate},
    morekeywords=[6]{do, last, first, try, idtac, repeat},
    morecomment=[s]{(*}{*)},
    showstringspaces=false,
    morestring=[b]",
    morestring=[d]’,
    tabsize=3,
    extendedchars=true,
    sensitive=true,
    breaklines=false,
    basicstyle=\small,
    captionpos=b,
    columns=[l]flexible,
    identifierstyle={\ttfamily\color{black}},
    keywordstyle=[1]{\ttfamily\color{dkviolet}},
    keywordstyle=[2]{\ttfamily\color{dkgreen}},
    keywordstyle=[3]{\ttfamily\color{ltblue}},
    keywordstyle=[4]{\ttfamily\color{dkred}},
    keywordstyle=[5]{\ttfamily\color{dkred}},
    keywordstyle=[6]{\ttfamily\color{dkred}},
    stringstyle=\ttfamily,
    commentstyle={\ttfamily\color{dkgreen}},
    literate=
    {\ :\ }{{\code{:}\,}}1
    {\\forall}{{\color{dkgreen}{$\forall\;$}}}1
    {\\exists}{{$\exists\;$}}1
    {<-}{{$\leftarrow\;$}}1
    {->}{{$\,\rightarrow\,$}}1
    {\ ->\ }{{$\;\rightarrow\;$}}1
    {<->}{{\,$\longleftrightarrow\;$\,}}1
    {<==}{{$\leq\;$}}1
    {\#}{{$^\star$}}1
    {\\o}{{$\circ\;$}}1
    {\@}{{$\cdot$}}1
    {\/\\}{{$\wedge\;$}}1
    {\\\/}{{$\vee\;$}}1
    {++}{{\,\raisebox{0.1em}{\tiny ++}\,}}1
    {\ ++\ }{{\;\raisebox{0.1em}{\tiny ++}\;}}1
    {:<}{{\,\code{:\!\raisebox{-0.04em}{\smaller{<}}}\,}}1
    {\ :<\ }{{\,\code{:\!\raisebox{-0.04em}{\smaller{<}}}\,}}1
    {≠}{{\,$\not=$\,}}1
    {~}{{$\,\sim\,$}}1
    {\@\@}{{$@$}}1
    {\\mapsto}{{$\mapsto\;$}}1
    {\\hline}{{\rule{\linewidth}{0.5pt}}}1
    {α}{{$\alpha$}}1
    {=}{{$\eqv$}}1
    {=α}{{$\eqv[\alpha]\hspace{-0.5pt}$}}1
    {=αd}{{$\eqv[\alpha d]$}}1
    {=vα}{{$\eqv[v\hspace{-0.5pt}\alpha]\hspace{-0.5pt}$}}1
    {=v?}{{$\eqv[v\hspace{-0.5pt}?]\hspace{-0.5pt}$}}1
    {\ =\ }{{$\eqv$}}1
    {\ =α\ }{{$\hspace{0.5pt}\eqv[\alpha]\hspace{-0.5pt}$}}1
    {\ =αd\ }{{$\;\eqv[\alpha d]\;$}}1
    {\ =vα\ }{{$\hspace{0.5pt}\eqv[v\hspace{-0.5pt}\alpha]\hspace{-0.5pt}$}}1
    {\ =v?\ }{{$\hspace{0.5pt}\eqv[v?]$}}1
    {❬}{{$\left<\right.\mkern-2mu$}}1
    {❭}{{$\mkern-2mu\left.\right>$}}1
    {❬\ }{{$\left<\right.$}}1
    {\ ❭}{{$\left.\right>$}}1
    {\ ;}{{;\,}}1
    {;\ }{{;\,}}1
    {\ ;\ }{{;\,}}1
    {x1}{{\texttt{x}$_1$\;\!\!}}2
    {x1\ }{{\texttt{x}$_1$\;}}2
    {x2}{{\texttt{x}$_2$\;\!\!}}2
    {x2\ }{{\texttt{x}$_2$\;}}2
    {y1}{{\texttt{y}$_1$\;\!\!}}2
    {y1\ }{{\texttt{y}$_1$\;}}2
    {y2}{{\texttt{y}$_2$\;\!\!}}2
    {y2\ }{{\texttt{y}$_2$\;}}2
    {xs1}{{\texttt{xs}$_1$\;\!\!}}3
    {xs1\ }{{\texttt{xs}$_1$\;}}3
    {xs2}{{\texttt{xs}$_2$\;\!\!}}3
    {xs2\ }{{\texttt{xs}$_2$\;}}3
    {xs3}{{\texttt{xs}$_3$\;\!\!}}3
    {xs3\ }{{\texttt{xs}$_3$\;}}3
    {ys1}{{\texttt{ys}$_1$\;\!\!}}3
    {ys1\ }{{\texttt{ys}$_1$\;}}3
    {ys2}{{\texttt{ys}$_2$\;\!\!}}3
    {ys2\ }{{\texttt{ys}$_2$\;}}3
    {ys3}{{\texttt{ys}$_3$\;\!\!}}3
    {ys3\ }{{\texttt{ys}$_3$\;}}3
    {≠}{{\,$\not=$\,}}1
    {⎡}{{$\left[\right.$}}1
    {⎤}{{$\left.\right]$}}1
    {⎤\ ~}{{$\left.\right]\!\sim\ \!$}}2
    {→}{{$\to$}}1
    {\ →\ }{{$\,\to\,$}}1
    {≠}{{\code{≠ }}}1
    {∉}{{$\not\in\ $}}1
    {∈}{{$\in$}}1
    {\ ∈\ }{{$\;\in\;$}}1
    {∈?}{{$\in_?$}}2
    {\ ∈?\ }{{$\;\in_?\;$}}2
    {≠}{{$\not=\ $}}1
    {∀}{{$\forall$}}1
    {∀\ }{{$\forall\,$}}1
    {λ}{{$\lambda\ $}}1
    {⊆}{{$\subseteq\ $}}1
    {,}{{,\;}}1
    {--}{{\,$\setminus$\,}}1
    {\ --\ }{{\,$\setminus$\,}}1
    {=>}{{$\Rightarrow\;$}}1
    {==}{{\code{==}\;}}1
    {==>}{{\code{==>}\;}}1
    {:=}{{\code{:=}}}1
    {\ :=\ }{{\code{:=} }}1
}[keywords,comments,strings]

\keywords{lambda calculus, alpha-equivalence, formalization, Rocq Prover}

\begin{document}

\title{A simple formalization of alpha-equivalence}

\author[K.~Apinis]{Kalmer Apinis\lmcsorcid{0009-0006-2395-6584}}
\author[D.~Ahman]{Danel Ahman\lmcsorcid{0000-0001-6595-2756}}

\address{
	Institute of Computer Science, University of Tartu, Estonia}
\email{kalmer.apinis@ut.ee, danel.ahman@ut.ee}

\begin{abstract}
  While teaching untyped $\lambda$-calculus to undergraduate students, we were left wondering why $\alpha$-equivalence is not directly inductively defined.
  In this paper, we demonstrate that this is indeed feasible.
  Specifically, we provide a grounded, inductive definition for $\alpha$-equivalence and show that it conforms to the specification provided in the literature.
  The work presented in this paper is fully formalized in the Rocq Prover.
\end{abstract}

\maketitle

\section{Introduction}
\label{s:introduction}

ACM's Computer Science Curricula 2023~\cite{kumar2024computer} cites $\lambda$-calculus as a canonical formalism for reasoning about functions, computation, and programming language semantics, reflecting its established role in computer science education.
Interesting $\lambda$-calculus properties can be stated concisely, such that they are suitable for teaching to undergraduate students. However, their formal proofs have caveats --- they are not all easy to directly prove to undergraduate students.
To address this issue, we construct a grounded, inductive definition of $\alpha$-equivalence such that many of its known properties are directly provable without involving more complicated mathematical structures.

As proposed by \citet{church1941calculi}, $\lambda$-calculus (originally called the \emph{calculus of $\lambda$-K-conversions}) defines three conversion rules.
The first of these rules says that we are free to rename any local variable $x$ to any variable $y$ which does not occur in the term --- establishing the intuitive notion that "the names of local variables do not matter".
This rule was later (already in \cite{curry1958combinatory}) called the $\alpha$-rule.
Nevertheless, for Church, the first rule is still a conversion --- no explicit notion is introduced that such a transformation produces an equivalent term in any general sense.
Moreover, the conversion operation only defines a choice --- we are not \emph{explicitly} compelled to perform any specific transformation.

However, as defined by Church, the $\alpha$-rule can be used to convert the term to meet the preconditions for the second rule.
The second conversion rule states that we can replace a function call with the body of the function, where the actual argument replaces the formal parameter, provided that local variables are distinct from parameters and global variables.
This rule was later named the $\beta$-rule.

Subsequently, $\lambda$-terms are viewed as representatives of equivalence classes induced by the $\alpha$-rule. Then, conversion using the $\alpha$-rule is either performed implicitly (e.g., following \citet{barendregt1984lambda}) or as part of the substitution while performing the conversion using the $\beta$-rule (e.g., \cite{hindley1986introduction}).
To identify $\lambda$-terms with the equivalence classes they denote, one can forgo variable names altogether and work with de Bruijn indices \cite{de1972lambda}.

Using the alternative approach with de Bruijn indices instead of variable names can effectively reduce $\alpha$-equivalence problems to equality problems.
However, this only changes rather than removes the proof obligations when we want to reason about terms with variable names. Then, named terms must be translated and named operations must be related to operations on their translations. Then, while some final proofs may become very short, the required correspondence layer is non-trivial \cite{gordon1993mechanisation}. We compare these alternatives in Section~\ref{s:debruijn}.

Implicit application of the $\alpha$-rule was popularised by \citet{barendregt1984lambda} --- meaning that the precondition for the $\beta$-rule and issues related to $\alpha$-equivalence are effectively ignored.
This, however, is unwarranted for undergraduate students who must first learn the basics of $\lambda$-calculus.
Moreover, omitting the tricky parts can be a drawback in a formal development.

A concise method for proving properties of $\lambda$-terms was introduced by \citet{gordon1996five}, by defining \emph{alpha}-structural principles that provide an induction principle for $\alpha$-equivalence classes.
This has led to interesting results \cite{pitts2006alpha,aydemir2007nominal, urban2008nominal, copello2016alpha}.

Explicit use of the $\alpha$-rule while performing a substitution is already mentioned by \citet{curry1958combinatory}.
In the literature, some definitions of $\alpha$-equivalence define it as just the equivalence relation (reflexive, symmetric, transitive relation) that also contains the $\alpha$-rule~\cite{hindley1986introduction, hankin1994lambda}, while others define it as the subcongruence one gets from $\beta$-reduction when removing the $\beta$-rule~\cite{mckinna1999some}, meaning that $\alpha$–equivalent terms are automatically also $\beta$–equivalent.
Formal (nominal) definitions sometimes also employ a swap operation, which swaps two given variable names (atoms) with each other in a given term~\cite{pitts2003nominal,urban2005formal,aydemir2007nominal,copello2016alpha} or
use a disjoint set of parameters with textual substitution~\cite{pollack1993closure,mckinna1993pts}.
However, note that such a swap/substitution operation generally does not follow variable scopes.
Thus, in some cases, one might prefer not to introduce additional technical operations that do not follow the usual semantics of $\lambda$-terms---as also noted by Pollack~\cite{pollack1995theory}.
While others have also considered and proposed more direct definitions of $\alpha$-equivalence \cite{SEBauer, BauerBlog}, we have not found any published formalizations of such definitions with accompanying proofs.

Thus, the question arises whether it is reasonable to formalize $\alpha$-equivalence of untyped $\lambda$-terms in the style of \citet{hindley1986introduction} and \citet{curry1958combinatory}.
We answer this question positively using the Rocq Prover (v.~9.1.1, together with the Equations plugin v.~1.3.1) \cite{RocqProver,sozeau2019equations}, and present the resulting formalization in a form intended to support instructors and students when $\alpha$–equivalence and its consequences are first taught.

\paragraph{Structure of the paper.}
We begin by recalling the definition of untyped $\lambda$-calculus terms in Section~\ref{s:lam}. In Section~\ref{s:alpha},
we give $\alpha$-equivalence an inductive definition
via a set of straightforward inference rules.
In Section~\ref{s:aprop}, we show that the proposed inference rules not only define an equivalence relation but also satisfy properties in the literature.
In addition, we show that $\alpha$-equivalence satisfies structural rules of equational logic.
We then define a specification for substitution in Section~\ref{s:ssubst} and show, in Section~\ref{s:sprop}, that it preserves $\alpha$-equivalence as expected.
In Section~\ref{s:subst_impl}, we show that a practical capture-avoiding implementation of substitution agrees with our specification relation.
In Section~\ref{s:adec}, we provide a decision procedure for checking $\alpha$-equivalence.
In Section~\ref{s:bvc}, we consider the variable convention~\cite{barendregt1984lambda}, which allows us to use non-capture-avoiding substitution safely --- this is used to prove the substitution lemma~\cite{barendregt1984lambda} in Section~\ref{s:subst_lemma}.
In Section~\ref{s:debruijn}, we compare our direct approach with approaches using de Bruijn indices.
We conclude in Section~\ref{s:conc}. The accompanying Rocq code without proofs is given in Appendix~\ref{a:rocq-formalization}, with the full formalization and proofs available at \url{https://github.com/kalmera/lambda-simple-alpha}. In the paper, a small \rocqrepo{icon} links to the corresponding declaration or proof on GitHub.

\paragraph{Formalization}
Our Rocq formalization is organized around four goals. First, we show that the
proposed definition of $\alpha$-equivalence is an equivalence and a congruence
(Section~\ref{s:aprop}). Second, we show that it interacts correctly with a
relational specification of capture-avoiding substitution
(Section~\ref{s:sprop}). Third, we verify executable implementations of
substitution, $\alpha$-equivalence checking, and freshening against their
specifications (Section~\ref{s:subst_impl}-\ref{s:bvc}). Finally, we use
freshening and the variable convention to prove the substitution lemma
(Section~\ref{s:subst_lemma}). The auxiliary results we develop serve these
goals: the context-manipulation and free-variable properties support the
generalized substitution theorem, while the variable-convention and freshening
properties reduce the final result to a lemma about simple substitution.

\section{Untyped $\lambda$-calculus\label{s:lam}}

Let $V$ be a countable set of variables, ranged over by $x, y, z, \ldots$. The
grammar
\[
\Lambda ::= V \;|\; (\Lambda\ \Lambda) \;|\; (\lambda V .\; \Lambda)
\]
defines the terms of the untyped $\lambda$-calculus \cite{church1941calculi}. A
term can be a variable, an application of two terms, or a $\lambda$-abstraction
consisting of a parameter variable and a term.
The convention is to avoid parentheses by assuming that application is left-associative and abstraction binds as much on the right as possible. For example,
the term $\lambda x.\ \lambda y.\ x\ y\ z$ is the same as $(\lambda x.\ (\lambda y.\ ((x\ y)\ z)))$.
At the top level, a term $y\in V$ denotes the global variable $y$.
An abstraction $(\lambda x.\ t)$, where $t\in\Lambda$, denotes a function whose parameter is the variable $x$ and body is the term $t$.
In the function body $t$, we can refer to the function's parameter using the variable $x$.
For example, $(\lambda x.\ x)$ is the identity function, whereas $(\lambda x.\ y)$ is the constant function that discards its argument and returns the global variable $y$.
Application $(f\ t)$, where $f, t\in\Lambda$, denotes applying the function $f$ to the argument $t$.
For example, $(\lambda x.\ x)\ z$ denotes the result of applying the identity function to the global variable $z$. The result is the global variable $z$.

Our Rocq implementation of the untyped $\lambda$-calculus is parameterized over a type \lstinline|V| of variables. For each pair of variables \lstinline|x, y : V|, our Rocq code is also parameterized over a function \lstinline|x =v? y| for checking variable equality. Using the equality check, we define a decision procedure \lstinline|x  ∈? xs| to check membership of \lstinline|x| in the variable list \lstinline|xs|. Furthermore, we assume that we can generate fresh variables using the function \lstinline|fresh| such that \lstinline|∀ l, fresh l ∉ l|.
We then define the terms of untyped $\lambda$-calculus as the following inductive \rocqsource{LamDirect.v}{437}{datatype} with three constructors:
\begin{lstlisting}
Inductive Lam : Type :=
  | Var : V -> Lam
  | App : Lam -> Lam -> Lam
  | Abs : V -> Lam -> Lam.
\end{lstlisting}
For example, the previously introduced term $\lambda x.\ \lambda y.\ x\ y\ z$ of the untyped $\lambda$-calculus is encoded in Rocq as \lstinline|Abs $x$ (Abs $y$ (App (App (Var $x$) (Var $y$)) (Var $z$)))| of type \lstinline|Lam|.

In untyped $\lambda$-calculus, local variables are called \emph{bound} variables as the use of a variable $x$ is said to be bound to an enclosing abstraction $(\lambda x.\ \ldots x \ldots)$, where it will get its value. All variables that occur in a term but are not bound are called \emph{free}. For any term $t\in\Lambda$, we denote the set of all occurring variables as $\Vars(t)$ and the set of free variables as $\FV(t)$. For example, $\Vars(\lambda x.\ x\ (\lambda y.\ z)) = \{x, y, z\}$ and $\FV(\lambda x.\ x\ (\lambda y.\ z)) = \{z\}$.

\section{Definition of $\alpha$-equivalence\label{s:alpha}}

The intuitive idea that "the names of local variables do not matter" is captured by $\alpha$-equivalence \cite{church1941calculi, curry1958combinatory}. For example, $(\lambda x.\ \lambda y.\ y)$ encodes the function that takes two arguments and returns the second one, as do $(\lambda a. \ \lambda b.\ b)$ and $(\lambda y.\ \lambda x.\ x)$. We denote this by writing $(\lambda x.\ \lambda y.\ y) =_\alpha (\lambda a. \ \lambda b.\ b) =_\alpha (\lambda y.\ \lambda x.\ x)$.
Now, if we want to consider the terms $\alpha$-equivalent, $y$ in the context $(\lambda x.\ \lambda y.\ \_)$ must be $\alpha$-equivalent to $b$ in the context $(\lambda a. \ \lambda b.\ \_)$.
To state this formally, we do not need to consider the full context of the respective subexpression, but only the respective abstraction variables outside the subexpression.
Thus, we might write the $\alpha$-equivalence of these subexpressions as $\ang{[x,y]; y} =_\alpha \ang{[a,b]; b}$, where we explicitly keep track of the respective abstraction variables.\footnote{Throughout this paper, we use so-called \emph{snoc}-lists (with type \lstinline|slist: Type -> Type|). We do so because we want to keep variable names in the same order in a list as they appear in the term, and we regularly need to append elements and take elements away from the end of the list. We write $\textit{xs} \snocop x$ for appending $x$ to the right end of $\textit{xs}$. Thus, any list $\textit{xs}$ is either empty or has the form $\textit{xs'} \snocop x$; for example, $[x,y] = [] \snocop x \snocop y$.}

\begin{figure}
\begin{align*}
  \inferrule[GlobalVA]
    { }
    {\ang{[]; x} \eqv[v\alpha] \ang{[]; x}}&&
  \inferrule[HereVA]
    { }
    {\ang{\textit{xs} \snocop x; x} \eqv[v\alpha] \ang{\textit{ys} \snocop y; y}}&&
  \inferrule[ThereVA]
    {\ang{\textit{xs}; a} \eqv[v\alpha] \ang{\textit{ys}; b} \\ x\neq a \\ y\neq b}
    {\ang{\textit{xs} \snocop x; a} \eqv[v\alpha] \ang{\textit{ys} \snocop y; b}}
  \end{align*}

  \medskip

  \begin{align*}
  \inferrule[VarA]
    {\ang{\textit{xs}; x} \eqv[v\alpha] \ang{\textit{ys}; y} }
    {\ang{\textit{xs}; x} \eqv[\alpha] \ang{\textit{ys}; y}}&&
  \inferrule[AppA]
    {\ang{\textit{xs}; f} \eqv[\alpha] \ang{\textit{ys}; g} \\ \ang{\textit{xs}; t} \eqv[\alpha] \ang{\textit{ys}; u}}
    {\ang{\textit{xs}; f\ t} \eqv[\alpha] \ang{\textit{ys}; g\ u}} &&
  \inferrule[AbsA]
    {\ang{\textit{xs} \snocop x; t} \eqv[\alpha] \ang{\textit{ys} \snocop y; u}}
    {\ang{\textit{xs}; \lambda x.\ t} \eqv[\alpha] \ang{\textit{ys}; \lambda y.\ u}}
  \end{align*}
  \caption{Inference rules for $\alpha$-equivalence.\label{f:a}}
\end{figure}
\noindent
In detail, we define $\alpha$-equivalence using two inductively defined binary relations, one for variables and one for terms, written $\ang{\textit{xs}; x} \eqv[v\alpha] \ang{\textit{ys}; y}$ and $\ang{\textit{xs}; t} \eqv[\alpha] \ang{\textit{ys}; u}$, both indexed by two abstraction variable contexts $\textit{xs}$ and $\textit{ys}$. These two relations are defined in Figure~\ref{f:a}.

Let us first look at $\alpha$-equivalence of variables.
For all $x\in V$, $\ang{[]; x} \eqv[v\alpha] \ang{[]; x}$, i.e., in the case of empty contexts, a variable can only be related to itself.
Next, for all variables $x, y\in V$ and variable lists $\textit{xs}$ and $\textit{ys}$, $\ang{\textit{xs} \snocop x; x} \eqv[v\alpha] \ang{\textit{ys} \snocop y; y}$, i.e., if neither context is empty, the rightmost entries of the contexts are related. Finally, for all variables $x, y, a, b\in V$ and variable lists $\textit{xs}, \textit{ys}$, if we have $\ang{\textit{xs}; a} \eqv[v\alpha] \ang{\textit{ys}; b}$, $x\neq a$, and $y\neq b$, then $\ang{\textit{xs} \snocop x; a} \eqv[v\alpha] \ang{\textit{ys} \snocop y; b}$, i.e., variables $a$ and $b$ at lower positions in the context can be related only if neither $a$ nor $b$ appears above in its respective context. This relation can be formalized as the following \rocqsource{LamDirect.v}{486}{Rocq definition}:

\begin{lstlisting}
Inductive VarAlpha : slist V -> slist V -> V -> V -> Prop :=
  | GlobalVA x : ❬[]; x❭ =vα ❬[]; x❭
  | HereVA xs ys x y : ❬xs :< x; x❭ =vα ❬ys :< y; y❭
  | ThereVA xs ys x y a b :
      x ≠ a -> y ≠ b -> ❬xs; x❭ =vα ❬ys; y❭ -> ❬xs :< a; x❭ =vα ❬ys :< b; y❭
where "❬ xs ; x ❭ =vα ❬ ys ; y ❭" := (VarAlpha xs ys x y).
\end{lstlisting}

\noindent
Next, we look at $\alpha$-equivalence of terms.
Variable terms $x, y\in V$ in respective contexts $\textit{xs}, \textit{ys}$ are $\alpha$-equivalent (i.e., $\ang{\textit{xs}; x} \eqv[\alpha] \ang{\textit{ys}; y}$) if they are $\alpha$-equivalent as variables, i.e., if $\ang{\textit{xs};x} \eqv[v\alpha] \ang{\textit{ys}; y}$. Applications $(f\ t)\in\Lambda$ and $(g\ u)\in\Lambda$ in respective contexts $\textit{xs}, \textit{ys}$ are $\alpha$-equivalent (i.e., $\ang{\textit{xs}; (f\ t)} \eqv[\alpha] \ang{\textit{ys}; (g\ u)}$) if their function and argument components are $\alpha$-equivalent, i.e., when $\ang{\textit{xs};f} \eqv[\alpha] \ang{\textit{ys}; g}$ and $\ang{\textit{xs};t} \eqv[\alpha] \ang{\textit{ys}; u}$. Abstractions $(\lambda x.\ t)\in\Lambda$ and $(\lambda y.\ u)\in\Lambda$ in respective contexts $\textit{xs}, \textit{ys}$ are $\alpha$-equivalent (i.e., $\ang{\textit{xs}; (\lambda x.\ t)} \eqv[\alpha] \ang{\textit{ys}; (\lambda y.\ u)}$) if terms $t$ and $u$ are $\alpha$-equivalent in extended contexts $\textit{xs} \snocop x$ and $\textit{ys} \snocop y$, respectively, i.e., when $\ang{\textit{xs} \snocop x; t} \eqv[\alpha] \ang{\textit{ys} \snocop y; u}$. This relation between terms can be formalized as the following \rocqsource{LamDirect.v}{495}{Rocq definition}:

\begin{lstlisting}
Inductive Alpha : slist V -> slist V -> Lam -> Lam -> Prop :=
  | VarA xs ys x y :
      ❬xs; x❭ =vα ❬ys; y❭ -> ❬xs; Var x❭ =α ❬ys; Var y❭
  | AppA xs ys f g t u :
      ❬xs; f❭ =α ❬ys; g❭ -> ❬xs; t❭ =α ❬ys; u❭ -> ❬xs; App f t❭ =α ❬ys; App g u❭
  | AbsA xs ys x y t u :
      ❬xs :< x; t❭ =α ❬ys :< y; u❭ -> ❬xs; Abs x t❭ =α ❬ys; Abs y u❭
where "❬ xs ; t ❭ =α ❬ ys ; u ❭" := (Alpha xs ys t u).
\end{lstlisting}

\noindent
We say that two terms $t$ and $u$ are $\alpha$-equivalent, written $t \eqv[\alpha] u$, if they are $\alpha$-equivalent in the empty context, i.e., $t \eqv[\alpha] u \iff \ang{[];t} \eqv[\alpha] \ang{[]; u}$. Thus, using these definitions, it is easy to prove that $(\lambda x.\ \lambda y.\ y) \eqv[\alpha] (\lambda a.\ \lambda b.\ b)$ because we can derive $\ang{[x,y];y} =_\alpha \ang{[a,b];b}$. Additionally, see Example~\ref{ex:shadow} for a proof that $\lambda x.\ x\ (\lambda x.\ y\ x) \eqv[\alpha] \lambda a.\ a\ (\lambda b.\ y\ b)$ also holds.

\begin{exa}\label{ex:shadow}
  Let us investigate whether $\lambda x.\ x\ (\lambda x.\ y\ x) \eqv[\alpha] \lambda a.\ a\ (\lambda b.\ y\ b)$ holds. To show that these two open $\lambda$-terms are $\alpha$-equivalent, we construct the following derivation tree:

  \[
    \inferrule*
      {\inferrule*
        {\inferrule*
          {\inferrule*
            {\phantom{.}}
            {\ang{[x]; x} \eqv[v\alpha] \ang{[a]; a}}}
          {\ang{[x]; x} \eqv[\alpha] \ang{[a]; a}}
         \\
         \inferrule*
          {\inferrule*
            {\inferrule*
              {\inferrule*
                {\inferrule*
                  {\inferrule*
                    { }
                    {\ang{[]; y} \eqv[v\alpha] \ang{[]; y}}}
                  {\ang{[x]; y} \eqv[v\alpha] \ang{[a]; y}}}
                {\ang{[x,x]; y} \eqv[v\alpha] \ang{[a,b]; y}}}
              {\ang{[x,x]; y} \eqv[\alpha] \ang{[a,b]; y}}
              \\
             \inferrule*
              {\inferrule*
                {\phantom{=}}
                {\ang{[x,x]; x} \eqv[v\alpha] \ang{[a,b]; b}}}
              {\ang{[x,x]; x} \eqv[\alpha] \ang{[a,b]; b}}}
            {\ang{[x,x]; y\ x} \eqv[\alpha] \ang{[a,b]; y\ b}}}
          {\ang{[x]; \lambda x.\ y\ x} \eqv[\alpha] \ang{[a]; \lambda b.\ y\ b}}}
        {\ang{[x]; x\ (\lambda x.\ y\ x)} \eqv[\alpha] \ang{[a]; a\ (\lambda b.\ y\ b)}}}
      {\ang{[]; \lambda x.\ x\ (\lambda x.\ y\ x)} \eqv[\alpha] \ang{[]; \lambda a.\ a\ (\lambda b.\ y\ b)}}
  \]

\end{exa}

\noindent
Our definition of $\alpha$-equivalence as a relation on $\lambda$-terms indexed by contexts is similar to the one proposed by \citet{SEBauer}, but with the difference that we represent contexts as a pair of lists, rather than a list of pairs. We found that this significantly simplifies the formalization.
Our definition also bears a resemblance to the inductive relation used by
McKinna and Pollack~\cite{mckinna1993pts,pollack1993closure,mckinna1999some}, who specify
$\alpha$-equivalence for binders by replacing the bound variables in the left-
and right-hand terms with the same (suitably fresh) parameter. 
Proofs of \emph{their} formal definitions are described in Pollacks PhD thesis~\cite{pollack1995theory}.
Our relation
$\ang{\textit{xs}; t} \eqv[\alpha] \ang{\textit{ys}; u}$ can be seen as a
linearized variant of their idea: we do not insist that bound variables under
corresponding binders have the same names in the left- and right-hand terms;
instead, their positions in the contexts identify them.

\section{Properties of $\alpha$-equivalence\label{s:aprop}}

We begin with some standard properties as a sanity check to ensure that the $\alpha$–equivalence relation we defined in the previous section makes sense.
First, the defined relation is an equivalence relation, i.e., it is \emph{reflexive}, \emph{symmetric}, and \emph{transitive}.
Further, contextual $\alpha$-equivalence preserves free variables after the
variables represented by each context are removed. In particular, ordinarily
$\alpha$-equivalent terms, whose contexts are empty, have equal free variables.
As our mathematical notation is similar to the formulation in Rocq, from now on, we will only show
the (slightly more verbose) Rocq notation for such properties.
We can prove the following \rocqsource{LamDirect.v}{118}{properties}:
\begin{lstlisting}
Property AlphaRefl : ∀ t xs,               ❬xs; t❭ =α ❬xs; t❭.
Property AlphaSym : ∀ t u xs ys,           ❬xs; t❭ =α ❬ys; u❭ -> ❬ys; u❭ =α ❬xs; t❭.
Property AlphaTrans: ∀ t u v xs ys zs,❬xs; t❭=α❬ys; u❭ -> ❬ys; u❭=α❬zs; v❭ -> ❬xs; t❭=α❬zs; v❭.
Property AlphaFV : ∀ t u xs ys,            ❬xs; t❭ =α ❬ys; u❭ -> FV t -- xs = FV u -- ys.
\end{lstlisting}

\noindent
Furthermore, our $\alpha$-equivalence definition satisfies the 'logical' axioms of \citet{barendregt1984lambda}, i.e., we can extend $\alpha$-equivalent terms with equal components. In other words, this property shows that $\alpha$-equivalence (\!$\eqv[\alpha]$\!) is a congruence.
We proved the \rocqsource{LamDirect.v}{129}{properties}:
\begin{lstlisting}
Property AlphaConvAppFun : ∀ (t u v : Lam), $\;\!\;$t =α u -> App t v =α App u v.
Property AlphaConvAppArg : ∀ (t u v : Lam), $\;\!\;$t =α u -> App v t =α App v u.
Property AlphaConvAbs : ∀ (t u : Lam) (x : V), t =α u -> Abs x t =α Abs x u.
\end{lstlisting}

\noindent
The remaining properties in this section---shadow removal, explicit treatment
of global variables, weakening, and exchange---provide the context-manipulation
tools needed for the generalized substitution theorem in
Section~\ref{s:sprop}.

Looking at Example~\ref{ex:shadow}, we note that context lists can contain duplicates. Furthermore, note that $\ang{[x,x]; x}$ would not be $\alpha$-equivalent to $\ang{[b,a]; b}$, as the rightmost $x$ in the list effectively \emph{shadows} the left $x$. This means that we are prohibited from searching for another $x$ in the context: the currently accessible $x$ must be related to the rightmost entry of the other context --- $a$ in our example. To this end, we prove the following \rocqsource{LamDirect.v}{584}{Rocq property}: we may remove shadowed variables (marked in blue) from the context.
\begin{lstlisting}
Property AlphaRemShadowed :
  ∀ t xs3 ys3 u x x' y y' xs2 xs1 ys2 ys1,
  (x = x' \/ y = y') ->
  length xs3 = length ys3 ->
  length xs2 = length ys2 ->
  ❬xs1++[$\color{blue}\texttt{x'}$]++xs2++[x]++xs3; t❭ =α ❬ys1++[$\color{blue}\texttt{y'}$]++ys2++[y]++ys3; u❭ ->
  ❬xs1++xs2++[x]++xs3; t❭ =α ❬ys1++ys2++[y]++ys3; u❭.
\end{lstlisting}

\noindent
Note that, in effect, the rules \textsc{GlobalVA} and \textsc{ThereVA} given in~Figure~\ref{f:a} together state that a variable $x$ in context $\textit{xs}$ is $\alpha$-equivalent to $x$ in context $\textit{ys}$ if $x$ is neither in $\textit{xs}$ nor in $\textit{ys}$, but only if the two contexts have the same length.
This kind of implicit equivalence of global variables can be made explicit by adding the variable $x$ to the left of both contexts.
As a result, we get the following \rocqsource{LamDirect.v}{719}{property}, showing that we can always prepend any list to or remove any list from both contexts, provided the contexts have the same length:

\begin{lstlisting}
Property AlphaCtxImplGlobs : ∀ t u xs ys zs,
  length xs = length ys $\;$->$\;\,$ ❬xs; t❭ =α ❬ys; u❭ <-> ❬zs++xs; t❭ =α ❬zs++ys; u❭.
\end{lstlisting}

\noindent
We next establish standard structural properties of equational logic (e.g.,
see \cite{jacobs1999categorical}). Weakening permits adding variables at the
same position in both contexts, provided that each added variable is either
absent from the corresponding term or shadowed by a later context entry. The
corresponding \rocqsource{LamDirect.v}{630}{property} is:
\begin{lstlisting}
Property AlphaCtxWeaken :
  ∀ t u xs xs' ys ys' x y,
  length xs = length ys ->
  ❬xs' ++ xs; t❭ =α ❬ys' ++ ys; u❭ ->
  (x ∉ FV t \/ x ∈ xs) /\ (y ∉ FV u \/ y ∈ ys) ->
  ❬xs'++[x]++xs; t❭ =α ❬ys'++[y]++ys; u❭.
\end{lstlisting}

\noindent
This property also gives a compact example of how proofs about the contextual
relation proceed. In the following proof, \lstinline|H| is the original
$\alpha$-equivalence derivation and \lstinline|Hx| and \lstinline|Hy| are the
two weakening conditions.
\begin{lstlisting}[literate=
  {+}{{\tiny$\mathbf{+}$}}1    
  {=>}{{$\Rightarrow\;$}}1
  {//}{{\code{//} }}1
  {//=}{{\code{//=} }}1
  {\ :=\ }{{\code{:=} }}1
  {\ :<\ }{{\,\code{:\!\raisebox{-0.04em}{\smaller{<}}}\,}}1,
  morekeywords={into}
]
Proof.  move=> t u xs xs' ys ys' x y hlen H [Hx Hy].
        dependent induction H; constructor; simp FV in Hx, Hy.
        +   eapply VarAlphaWeaken => //; cbn in Hx, Hy; tauto.
        +   rewrite in_app_iff in Hx. rewrite in_app_iff in Hy.
            apply: IHAlpha1 => //=; by tauto.
        +   rewrite in_app_iff in Hx. rewrite in_app_iff in Hy.
            apply: IHAlpha2 => //=; by tauto.
        +   rewrite filter_In in Hx. rewrite filter_In in Hy.
            apply IHAlpha with (xs := xs :< x0) (ys := ys :< y0) => //=; auto.
        ++  elim: (in_list_spec x [x0]) Hx => //= Q Hx; tauto.
        ++  elim: (in_list_spec y [y0]) Hy => //= Q Hy; tauto.
Qed.
\end{lstlisting}
The induction follows the derivation of contextual $\alpha$-equivalence. In
the variable case, \lstinline|VarAlphaWeaken| supplies the corresponding
variable-level result. An application has two subderivations; membership in
the free-variable list of the application is split between them, giving the
side condition for each induction hypothesis. For abstractions with parameters
\lstinline|x0| and \lstinline|y0|, these parameters are appended to the inner
contexts before applying the induction hypothesis. Unfolding the filtered
free-variable lists verifies that the inserted variables are either absent
from the bodies or shadowed by these parameters. Thus, the proof follows the
three constructors of \lstinline|Alpha|, with the variable-level weakening
lemma containing the only context-specific base case.

We also prove that the variable exchange rule of equational logic is admissible.
We can swap the order of two consecutive context variables on both sides of
$\alpha$-equivalence at the same position, given that the consecutive variables
are not equal.
This precondition ensures we cannot swap a shadowed variable into the scope (we
have already seen rules for adding and removing shadowed variables). The
corresponding \rocqsource{LamDirect.v}{843}{property} is:
\begin{lstlisting}
Property AlphaSwap :
  ∀ t u xs ys xs' ys' a x b y,
  length xs = length ys ->
  ❬xs'++[a; x]++xs; t❭ =α ❬ys'++[b; y]++ys; u❭ ->
  a ≠ x -> b ≠ y ->
  ❬xs'++[x; a]++xs; t❭ =α ❬ys'++[y; b]++ys; u❭.
\end{lstlisting}

\section{Specifying substitutions\label{s:ssubst}}

The standard way of defining computation in $\lambda$-calculus is through reduction. When computing the meaning of applying an argument to a function, the term $(\lambda x.\ t)\ u$ reduces to $t$ where the uses of the formal parameter $x$ are substituted with the argument $u$. What makes substitution complicated is its interaction with bound variables and $\alpha$-equivalence.

To begin with, in this section, we define the (relational) specification of how a correct substitution operator for the $\lambda$-calculus has to behave.
We then use it later to study and prove the correctness of concrete implementations of substitution.
Notationally, we write $t[x\to u] \sim v$ to specify that substituting $u$ for $x$ in $t$ yields $v$.
Further, for convenience, we write $t[x\to u]$ to specify that any substitution result $v$ where $t[x\to u] \sim v$ is suitable in that position.
The substitution specification is then formally defined as the following relation:
\begin{align*}
  x[y\to u] &\sim
  \begin{cases}
    u,& \text{if }x=y\\
    x,& \text{otherwise}
  \end{cases}\\
  (f\ t)[y\to u] &\sim (f[y\to u]\ t[y\to u])\\
  (\lambda x.\ t)[y\to u] &\sim
    \begin{cases}
      (\lambda x.\ t),
          & \text{if }x=y\\
      (\lambda x.\ t[y\to u]),
          & \text{if }x\not= y \land x\not\in \textsf{FV}(u)
    \end{cases}
\end{align*}

\noindent
If the term in which we are substituting $u$ for $y$ is a variable $x$, then replacing free occurrences of $y$ with $u$ results in $u$ if $x=y$, and in $x$ otherwise.
Substitution in an application term $(f\ t)$ is defined by recursively replacing free occurrences of $y$ with $u$ in both $f$ and $t$.
We have two cases for substitution in abstractions $(\lambda x.\ t)[y\to u]$. If the variables $x$ and $y$ are the same, then $y$ is bound, and there are no free variables $y$ to substitute. We can return the abstraction as is. If the variables are different and $x$ does not occur freely in $u$, we can recursively continue substitution in the body of the abstraction.
Note that the specification does \emph{not} define a total function, because it does not specify what to do when the abstraction parameter $x$ occurs freely in the term $u$ (the potential capture case).

The substitution relation can be formalized using the following inductive \rocqsource{LamDirect.v}{472}{definition}. We will discuss some properties of the defined relation in the next section.
\begin{lstlisting}
Inductive Subst (x : V) (u : Lam) : Lam -> Lam -> Prop :=
  | SubstVarEq : (Var x)⎡x→u⎤~u
  | SubstVarNeq y : y ≠ x -> (Var y)⎡x→u⎤~Var y
  | SubstApp f f' t t' : f⎡x→u⎤~f' -> t⎡x→u⎤~t' -> (App f t)⎡x→u⎤~(App f' t')
  | SubstAbsEq t : (Abs x t)⎡x→u⎤~(Abs x t)
  | SubstAbsNF y t t' :
      y ≠ x -> y ∉ FV u -> t⎡x→u⎤~t' -> (Abs y t)⎡x→u⎤~(Abs y t')
where "t ⎡ y → u ⎤ ~ v" := (Subst y u t v).
\end{lstlisting}

\noindent
A quick yet naive way of turning this relational specification into a total function is to ignore the mentioned side conditions in the abstraction case.
Such a (simple) substitution operator can be formalized as the following recursively defined \rocqsource{LamDirect.v}{1035}{function}:
\begin{lstlisting}
Equations ssubst (t : Lam) (x : V) (u : Lam) : Lam :=
ssubst (Var y) x u := if y =v? x then u else Var y;
ssubst (App f t) x u := App (ssubst f x u) (ssubst t x u);
ssubst (Abs y t) x u := if y =v? x then Abs y t else Abs y (ssubst t x u).
\end{lstlisting}

\noindent
However, as is well known, naively replacing free variable occurrences of $x$ as above can lead to \emph{variable capture} and break compositionality.
For instance, suppose we have the term
$(\lambda x.\ (\lambda z.\ x)\ y)\ z$.
Replacing occurrences of the free variable $x$ with $z$ in the subterm $(\lambda z.\ x)\ y$ yields $(\lambda z.\ z)\ y$, which reduces to $y$. On the other hand, $(\lambda z.\ x)\ y$ reduces to $x$, and thus the original term can be reduced to $(\lambda x.\ x)\ z$, which reduces to $z$. Thus, we found a violation of compositionality when using the simple definition of substitution.
As noted by \citet{curry1958combinatory}, we can avoid variable capture by using a capture-avoiding definition of substitution, where the key idea is to systematically rename local variables with fresh names as shown in the following example.
We will implement such a \emph{capture-avoiding} substitution function in Section~\ref{s:subst_impl}.

\begin{exa}\label{e:subst}
  Suppose we need to substitute $z$ for $x$ in $(\lambda z.\ x)\ y$. This is not directly possible, as $z\in\FV(z)$. The solution is to find an $\alpha$-equivalent term that does not use any variable from $\FV(z)=\{z\}$ as the abstraction parameter. One such term is $(\lambda q.\ x)\ y$, where $q=\texttt{fresh} \{z\}$.
  Then we can conclude that
  $((\lambda q.\ x)\ y)[x\to z] \sim ((\lambda q.\ z)\ y)$, or pick any other variable $w\neq z$ and conclude that $((\lambda w.\ x)\ y)[x\to z] \sim ((\lambda w.\ z)\ y)$. It is straightforward to prove that $(\lambda z.\ x)\ y \;\eqv[\alpha]\, (\lambda q.\ x)\ y \;\eqv[\alpha]\, (\lambda w.\ x)\ y$.
\end{exa}

\section{Properties of substitution\label{s:sprop}}

To further convince ourselves that our definition of $\alpha$-equivalence is reasonable, we check whether well-known properties of substitutions dealing with $\alpha$-equivalence are provable using our definitions.
In this section, we will look at some basic properties of substitution and then, in Section~\ref{s:subst_lemma}, we will use the basic properties to prove the substitution lemma~\cite{Barendregt92}.

As formulated by \citet{curry1958combinatory}, substituting a variable $y\notin\FV(t)$ for $x$ in the term $t$, yielding $t'$, must imply that $\lambda x.\ t \eqv[\alpha] \lambda y.\ t'$.
We note that this is a slight deviation from \citet{barendregt1984lambda}, who makes a stronger assumption, requiring that $y$ is fresh. With our definitions, this can be formalized as the following \rocqsource{LamDirect.v}{699}{property}:
\begin{lstlisting}
Property SubstFreshVarAlpha : ∀ t t' x y,
  y ∉ FV t -> t⎡x→Var y⎤~t' -> Abs x t =α Abs y t'.
\end{lstlisting}

\noindent
Another useful property of substitution is that substituting a variable $x$ that is not free in $t$ always yields the original term $t$. This can be formalized as the following \rocqsource{LamDirect.v}{748}{property}:
\begin{lstlisting}
Property AlphaSubstUnusedVar : ∀ t x u v,
  x ∉ FV t -> t⎡x→u⎤~v -> t = v.
\end{lstlisting}

\noindent
A central property of substitution, stated informally, is that substituting parameters of $\alpha$-equivalent abstraction terms with $\alpha$-equivalent arguments yields $\alpha$-equivalent terms.
Formally, this can be expressed with our definitions as the following \rocqsource{LamDirect.v}{948}{property}:
\begin{lstlisting}
Property SubstAlpha:
  ∀ t t' u u' v v' xs ys x y,
  ❬xs; u❭ =α ❬ys; u'❭ ->
  ❬xs; Abs x t❭ =α ❬ys; Abs y t'❭ ->
  t⎡x→u⎤~v ->
  t'⎡y→u'⎤~v' ->
  ❬xs; v❭ =α ❬ys; v'❭.
\end{lstlisting}

\noindent
Note that the condition \lstinline|❬xs; Abs x t❭ =α ❬ys; Abs y t'❭| in \lstinline|SubstAlpha| is equivalent to the condition \lstinline|❬xs :< x; t❭ =α ❬ys :< y; t'❭|, i.e., the parameter is at the end of the context.
However, we found that this form is unsuitable for proof by induction, as in the case where \lstinline|t| is an abstraction, we need to use the induction hypothesis in a setting where the parameter is no longer at the very end of the context.
Thus, to prove \lstinline|SubstAlpha|, we need to generalize the property to cover cases where the parameters are in the middle of the context instead of only on the right.
This change results in the property \lstinline|SubstAlphaMain| below, where
we now have the condition \lstinline{❬xs++[x]++xs'; t❭ =α ❬ys++[y]++ys'; t'❭}, where the parameters \lstinline|x| and \lstinline|y| can appear in the middle of the context.
We also have the conditions \lstinline{length xs' = length ys'}, \lstinline|x ∉ xs'|, and \lstinline|y ∉ ys'|, which ensure that \lstinline|x| and \lstinline|y| are related parameters for \lstinline|t| and \lstinline|t'|, respectively. The resulting \rocqsource{LamDirect.v}{873}{property} is:
\begin{lstlisting}
Property SubstAlphaMain:
  ∀ t t' u u' v v' xs ys x y xs' ys',
  length xs' = length ys' ->
  x ∉ xs' -> y ∉ ys' ->
  ❬xs++xs'; u❭ =α ❬ys++ys'; u'❭ ->
  ❬xs++[x]++xs'; t❭ =α ❬ys++[y]++ys'; t'❭ ->
  t⎡x→u⎤~v ->
  t'⎡y→u'⎤~v' ->
  ❬xs++xs'; v❭ =α ❬ys++ys'; v'❭.
\end{lstlisting}

\noindent
We prove \lstinline|SubstAlphaMain| by induction on the structure of
\lstinline|t|, inverting the two substitution derivations and the contextual
$\alpha$-equivalence derivation in each case. For a variable, inversion
determines whether it is the substituted variable; the variable
$\alpha$-equivalence derivation then determines the corresponding result on the
other side. For an application, both substitution derivations and the
$\alpha$-equivalence derivation split into their function and argument parts,
to which the two induction hypotheses apply.

The abstraction case contains the reason for the generalized statement. A
substitution derivation for an abstraction either stops because its parameter
is the substituted variable, or enters the body because the parameter is
different and fresh for the replacement. Inverting the two substitution
derivations therefore gives four cases.
\begin{enumerate}
\item If both substitutions stop, the resulting abstractions are related after
  removing the now-shadowed context entries with
  \lstinline|AlphaRemShadowed|.
\item If the substitution stops on the left but enters the body on the right,
  \lstinline|AlphaNotFree| transfers the fact that the left substituted
  variable is not free in its abstraction to the right abstraction. Since the
  right parameter differs from the right substituted variable, the latter is
  not free in the right body. By \lstinline|AlphaSubstUnusedVar|, the right
  substitution therefore leaves that body unchanged, reducing this case to
  the preceding one.
\item The case where the substitution enters only on the left is symmetric.
\item If both substitutions enter their bodies, their abstraction parameters
  are appended to \lstinline|xs'| and \lstinline|ys'|. The replacement terms
  remain $\alpha$-equivalent in these extended contexts by
  \lstinline|AlphaCtxWeakenLeft|, using the freshness premises of the two
  substitution derivations. The induction hypothesis then relates the two
  substituted bodies, and the abstraction constructor concludes the case.
\end{enumerate}

\noindent
Using \lstinline|SubstAlphaMain|, we can then prove the more specific property \lstinline|SubstAlpha| stated above.
We conclude this section by showing additional structural rules standard in equational logic (e.g., see \cite{jacobs1999categorical}) that are corollaries of \lstinline|SubstAlphaMain|. First, we prove that the contraction rule is admissible, allowing two variables to be replaced by one. We get the following \rocqsource{LamDirect.v}{928}{corollary}:
\begin{lstlisting}
Corollary Contraction :
  ∀ t t' v v' xs ys xs' ys' x1 x2 y1 y2,
  length xs = length ys ->
  x1 ∉ xs -> y1 ∉ ys ->
  x2 ∉ ([x1]++xs) -> y2 ∉ ([y1]++ys) ->
  ❬xs'++[x2; x1]++xs; t❭ =α ❬ys'++[y2; y1]++ys; t'❭ ->
  t⎡x2→Var x1⎤~v ->
  t'⎡y2→Var y1⎤~v' ->
  ❬xs'++[x1]++xs; v❭ =α ❬ys'++[y1]++ys; v'❭.
\end{lstlisting}
Next, we can prove a substitutivity principle, which states that substituting $\alpha$-equivalent terms into a term yields $\alpha$-equivalent results. This is formalized as the following \rocqsource{LamDirect.v}{970}{corollary}:
\begin{lstlisting}
Corollary Substitutivity :
  ∀ t x u u' v v',
  u =α u' -> t⎡x→u⎤~v -> t⎡x→u'⎤~v' -> v =α v'.
\end{lstlisting}
Finally, we can also prove a general congruence rule, which states that substituting a term into $\alpha$-equivalent terms results in $\alpha$-equivalent terms. This is formalized as the following \rocqsource{LamDirect.v}{977}{corollary}:
\begin{lstlisting}
Corollary Congruence :
  ∀ t t' u v v' xs ys x y xs' ys',
  length xs' = length ys' ->
  x ∉ xs' -> y ∉ ys' ->
  ❬xs++xs'; u❭ =α ❬ys++ys'; u❭ ->
  ❬xs++[x]++xs'; t❭ =α ❬ys++[y]++ys'; t'❭ ->
  t⎡x→u⎤~v -> t'⎡y→u⎤~v' ->
  ❬xs++xs'; v❭ =α ❬ys++ys'; v'❭.
\end{lstlisting}

\section{A substitution function}\label{s:subst_impl}

The relational definition for substitution is convenient for proofs about substitutions. However, a relational definition is not directly suitable for computing a term and for use in a computational semantics.
Thus, we prefer a function that performs renaming and substitution in one pass, i.e., the function implementing capture-avoiding substitution.
We define this function by combining simple substitution (\lstinline|ssubst|) with renaming variables where needed.

First, to prove termination below, we define the \emph{height} of terms and show that simple substitution with a variable does not alter the height in the following \rocqsource{LamDirect.v}{249}{Rocq definition and property}:
\begin{lstlisting}
Equations height (t : Lam) : nat :=
height (Var _) := 1;
height (App f t) := 1 + max (height f) (height t);
height (Abs _ t) := 1 + height t.

Property ssubst_height : ∀ t x z, height (ssubst t x (Var z)) = height t.
\end{lstlisting}

\noindent
We now define the capture-avoiding substitution function. As its definition is not primitive recursive and requires a non-trivial termination proof in Rocq,
we will use well-founded recursion via the height of the term. The omitted termination proof is straightforward.
We use the Equations plugin for the following \rocqsource{LamDirect.v}{1049}{definition} of the \lstinline|subst| function:
\begin{lstlisting}
Equations? subst (t : Lam) (x : V) (u : Lam) : Lam by wf (height t) :=
subst (Var v) x u :=
  if v =v? x then u else Var v;

subst (App f t) x u :=
  App (subst f x u) (subst t x u);

subst (Abs y t) x u :=
  if y =v? x then
    Abs y t
  else if y ∈? FV u then
    let z   := fresh (Vars t ++ FV u ++ [y; x]) in
    let t'  := ssubst t y (Var z) in
    Abs z (subst t' x u)
  else
    Abs y (subst t x u).
\end{lstlisting}

\noindent
Note that in a practical implementation, instead of using the \lstinline|fresh| function, generating fresh variables could, for example, be implemented using a state monad modelling an incrementing counter, such that we would not need to manage explicit lists of variable names.

Finally, we prove that the definition of \lstinline|subst| conforms to the specification provided by the relational definition of substitution. Namely, for each term \lstinline|t|, there exists an $\alpha$-equivalent term \lstinline|t'| such that \lstinline|t'⎡y→u⎤~subst t y u|. In detail, we prove the following \rocqsource{LamDirect.v}{1290}{property}:

\begin{lstlisting}
Property SubstExists : ∀ t u x,
  { t' : Lam | t' =α t & t'⎡x→u⎤~subst t x u }.
\end{lstlisting}

\noindent
There is a technical note on the trusted base of properties of well-founded recursive functions such as \lstinline|subst| in Rocq.
The \emph{proofs} of such properties sometimes use unfolding equations generated for well-founded recursive functions and therefore depend on the \emph{functional extensionality} assumption \lstinline|functional_extensionality_dep|. This is the case for the property \lstinline|SubstExists| and several properties in Sections 9 and 10. Previously defined properties of $\alpha$-equivalence (e.g., \lstinline|SubstAlphaMain|) do not use well-founded recursive functions and do not require functional extensionality.

\section{Decision procedure for $\alpha$-equivalence\label{s:adec}}

In this section, we demonstrate that our proposed definition for $\alpha$-equivalence straightforwardly lends itself to a functional implementation. Specifically, if we have two concrete terms $t$ and $u$, we can check whether they are $\alpha$-equivalent using a decision procedure.

First, we implement a recursive \rocqsource{LamDirect.v}{1302}{function} \lstinline|decVarAlpha|, shown below, for deciding $\alpha$-equivalence of variables \lstinline|x| and \lstinline|y| in respective contexts \lstinline|xs| and \lstinline|ys|. The function performs pattern-matching on the pair of contexts. If both contexts are empty, then variables are $\alpha$-equivalent only if they are equal, as in rule \textsc{GlobalVA}. If both contexts are non-empty, we must check if variables are equal to their respective rightmost context variable. If both variables are equal to their respective rightmost context variable, then the variables are $\alpha$-equivalent, as in rule \textsc{HereVA}. If neither variable is equal to their respective rightmost context variable, we use recursion with smaller contexts, as in rule \textsc{ThereVA}. If precisely one variable is equal to its respective rightmost context variable or if precisely one context is empty, then variables are not $\alpha$-equivalent, as no rule is applicable, and we return \lstinline|false|.
\begin{lstlisting}
Equations decVarAlpha (xs ys : slist V) (x y : V) : bool :=
decVarAlpha (xs' :< x') (ys' :< y') x y :=
  match x =v? x', y =v? y' with
  | true, true => true
  | false, false => decVarAlpha xs' ys' x y
  | _, _ => false
  end;

decVarAlpha [] [] x y := x =v? y;
decVarAlpha _ _ _ _ := false.
\end{lstlisting}

\noindent
Using \lstinline|decVarAlpha|, we implement the decision procedure \lstinline|decAlpha|, shown below, for checking $\alpha$-equivalence of terms \lstinline|t| and \lstinline|u| in their respective contexts \lstinline|xs| and \lstinline|ys|. The recursive function starts by pattern matching on the terms \lstinline|t| and \lstinline|u|. If both terms are variables, then we use \lstinline|decVarAlpha|, as in rule \textsc{VarA}. If both terms are applications, as in rule \textsc{AppA}, we recursively check that both functions and arguments are respectively $\alpha$-equivalent. If both terms are abstractions, then we recurse with contexts extended by the respective abstraction variables, as in rule \textsc{AbsA}.
In cases where terms are not of the same kind, we return \lstinline|false| as no rule can apply. The corresponding \rocqsource{LamDirect.v}{1312}{definition} is:
\begin{lstlisting}
Equations decAlpha (xs ys : slist V) (t u : Lam) : bool :=
decAlpha xs ys (Var x) (Var y) := decVarAlpha xs ys x y;
decAlpha xs ys (App f t) (App g u) := decAlpha xs ys f g && decAlpha xs ys t u;
decAlpha xs ys (Abs x t) (Abs y u) := decAlpha (xs :< x) (ys :< y) t u;
decAlpha xs ys _ _ => false.
\end{lstlisting}

\noindent
We can then prove the following correctness theorem: the decision procedure returns \lstinline|true| precisely when $\alpha$-equivalence holds. The corresponding \rocqsource{LamDirect.v}{1336}{property} is:
\begin{lstlisting}
Property decAlphaSpec : ∀ (xs ys : slist V) (t u : Lam),
  decAlpha xs ys t u = true <-> ❬xs; t❭ =α ❬ys; u❭.
\end{lstlisting}
Thus, we can decide $\alpha$-equivalence by computation. Let us look at some examples.

\begin{exa}
If we need to prove $\alpha$-equivalence of
$(\lambda q.\ z)\ y \eqv[\alpha] (\lambda w.\ z)\ y$, we can use \lstinline|decAlphaSpec| to convert the proposition into a boolean check and then use the \lstinline|reflexivity| tactic to force Rocq into checking that the returned boolean is true. See the following \rocqsource{LamTest.v}{117}{example}, where we note that variable names must be quoted in our Rocq notation:
\begin{lstlisting}
Example test1: ((λ "q". "z") "y") =α ((λ "w". "z") "y").
Proof.
  apply decAlphaSpec. reflexivity.
Qed.
\end{lstlisting}
\end{exa}

\begin{exa}
In a negative case, where we need to prove $(\lambda z.\ z)\ y \,\not\hspace{-1.3ex}\eqv[\alpha] (\lambda q.\ z)\ y$, we first assume that the $\alpha$-equivalence holds, then use \lstinline|decAlphaSpec| to convert it into a boolean check, and finally use \lstinline|discriminate| to derive \lstinline|False| from \lstinline|false = true|, as in the following \rocqsource{LamTest.v}{120}{example}:
\begin{lstlisting}
Example test2: ~ ((λ "z". "z") "y" =α (λ "q". "z") "y").
Proof.
  intro C. apply decAlphaSpec in C. discriminate C.
Qed.
\end{lstlisting}
\end{exa}

\section{Variable Convention\label{s:bvc}}

\citet{barendregt1984lambda} states a convention for variables: if terms $M_1, \ldots, M_n$ occur in a certain mathematical context (e.g., in a definition or proof), then all bound variables in these terms are chosen to be different from free variables.
Under this convention, or more specifically when the bound variables in $t$ are different from the free variables in $v$, we are ensured that $t[x\to v] \sim \text{\lstinline|ssubst|}\ t \ x \ v$, i.e., the simple substitution does not exhibit variable capture.

We can formalize a variant of this convention using the following inductively defined predicate \lstinline|VC xs t|, which holds when variables from the list \lstinline|xs| are not used as abstraction variables in the term \lstinline|t|. We get Barendregt's formulation of the convention by requiring \lstinline|VC (FV($M_1$) ++$\,\ldots\,$++ FV($M_n$)) $M_i$|. The corresponding \rocqsource{LamDirect.v}{1370}{definition} is:
\begin{lstlisting}
Inductive VC (xs : slist V) : Lam -> Prop :=
  | VCVar x : VC xs (Var x)
  | VCApp f t : VC xs f -> VC xs t -> VC xs (App f t)
  | VCAbs x t : x ∉ xs -> VC xs t -> VC xs (Abs x t).
\end{lstlisting}

\noindent
Notice how the constructors \lstinline|VCVar| and \lstinline|VCApp| add no restrictions to the term \lstinline|t| and how the constructor \lstinline|Abs v t| corresponding to the abstraction case adds the requirement \lstinline|v ∉ xs|.
Thus, we have a precondition \lstinline|VC xs t| that ensures that capture-avoiding substitution \lstinline|subst t x v| is $\alpha$-equivalent to the non-capture-avoiding substitution \lstinline|ssubst t x v| from Section~\ref{s:ssubst}, provided that $\textsf{FV}(\texttt{v}) \subseteq \texttt{xs}$.
Formally, we can prove the following \rocqsource{LamDirect.v}{1460}{property}:
\begin{lstlisting}
Property vcEq :
  ∀ t x xs v, FV v ⊆ xs -> VC xs t -> subst t x v =α ssubst t x v.
\end{lstlisting}

\noindent
Unfortunately, when considering reductions of $\lambda$-terms, verifying adherence to the variable convention only at the beginning of a reduction sequence is insufficient. The action of substitution on terms can violate the \lstinline|VC| predicate, as illustrated by the next example.
\begin{exa}
  Consider evaluating the term $(\lambda f.\ f\ f)\ (\lambda x.\ \lambda y.\ x\ y)$. Note that the variable convention holds initially. Now, let us perform $\beta$-reduction steps.

  First, we substitute the argument $\lambda x.\ \lambda y.\ x\ y$ for parameter $f$ in the function body, noting that the substitution $(f\ f)[f \to \lambda x.\ \lambda y.\ x\ y]$ yields $(\lambda x.\ \lambda y.\ x\ y)\ (\lambda x.\ \lambda y.\ x\ y)$. Now, the function and argument share abstraction variables, but the variable convention still holds.

  Second, substituting $\lambda x.\ \lambda y.\ x\ y$ for $x$ in $(\lambda y.\ x\ y)$ yields $\lambda y.\ (\lambda x.\ \lambda y.\ x\ y)\ y$. However, in the next step, we need to substitute the argument $y$ for $x$ in $\lambda y.\ x\ y$, but for that step, the variable convention is violated as $y$ is used as a bound variable in $\lambda y.\ x\ y$. Simple substitution incorrectly yields $\lambda y.\ y\ y$, whereas capture-avoiding substitution yields, for example, $\lambda z.\ y\ z$.
\end{exa}

\noindent
Thus, when considering reduction sequences of $\lambda$-terms, we may need to periodically re-establish the \lstinline|VC| predicate to be able to use simple (non-capture-avoiding) substitutions safely. A natural solution for this is to define a function \lstinline|vcRefresh| such that \lstinline|vcRefresh t vs| returns an $\alpha$-equivalent term $t'$ satisfying \lstinline|VC vs t'|. We give the following \rocqsource{LamDirect.v}{1113}{definition}:
\begin{lstlisting}
Equations? vcRefresh (t : Lam) (xs : slist V) : Lam by wf (height t) :=
vcRefresh (Var x) _ :=
    Var x;

vcRefresh (App f u) xs :=
    App (vcRefresh f xs) (vcRefresh u xs);

vcRefresh (Abs y u) xs :=
    if y ∈? xs then
        let y' := fresh (Vars u ++ xs :< y) in
        let u' := ssubst u y (Var y') in
        Abs y' (vcRefresh u' xs)
    else
        Abs y (vcRefresh u xs).
\end{lstlisting}

\noindent
This function $\alpha$-converts its parameter \lstinline|t| such that no variable in \lstinline|vs| is used as a local variable.
We can prove that \lstinline|vcRefresh t vs| indeed returns a term $\alpha$-equivalent to \lstinline|t|, and that the \lstinline|VC| predicate holds for the returned term with respect to the list of (free) variables \lstinline|vs|. These are the following \rocqsource{LamDirect.v}{346}{properties}:
\begin{lstlisting}
Property AlphaVcRefresh : ∀ t xs, t =α vcRefresh t xs.
Property vcRefreshProp : ∀ t xs, VC xs (vcRefresh t xs).
\end{lstlisting}

\noindent
However, we also note that the variable convention does not always need to be re-established at every step because, for example, the simple substitution \lstinline|ssubst| does not increase the sets of bound or free variables. For instance, we can prove the following \rocqsource{LamDirect.v}{1386}{Rocq property}:
\begin{lstlisting}
Property VCssubst : ∀ t x xs u,
  VC xs u -> VC xs t -> VC xs (ssubst t x u).
\end{lstlisting}

\noindent
Thus, to enforce the variable convention, we need to replace each term $M_i$ with its $\alpha$-equivalent term \lstinline|$M'_i$ := vcRefresh $M_i$ (FV($M_1$) ++$\,\ldots\,$++ FV($M_n$))| at every step where we consider more terms or where the set of bound or free variables increases.
We note that \lstinline|vcRefresh| returns an $\alpha$-equivalent term; therefore, the original and refreshed terms have the same free variables, i.e., $\FV(M_i) = \FV(M'_i)$.

\section{A proof of the substitution lemma\label{s:subst_lemma}}

To conclude, we now combine the previous properties to prove the substitution lemma as stated by \citet{Barendregt92}.
The proof of this lemma exemplifies how we can sometimes extend properties of the simple substitution function \lstinline|ssubst| to capture-avoiding substitution.

Let $t_1, u_1, u_2\in\Lambda$ be $\lambda$-terms. Suppose $x\not= y$ and $x\notin \textsf{FV}(u_2)$. Then the relational \rocqsource{LamDirect.v}{1598}{\emph{substitution lemma}} says that substitution distributes over itself:
\[t_1[x\to u_1][y\to u_2] \;\eqv[\alpha]\; t_1[y\to u_2][x\to u_1[y\to u_2]] .\]
This can be shown to hold inductively on the substitution specification. However, a much more challenging task is to prove an equivalent \rocqsource{LamDirect.v}{1566}{functional substitution lemma} for the capture-avoiding substitution function:
\[\subst\ (\subst\ t_1\ x\ u_1)\ y\ u_2 \;\eqv[\alpha]\; \subst\ (\subst\ t_1\ y \ u_2)\ x \ (\subst\ u_1\ y\ u_2) .\]
One major obstacle to a direct inductive proof (using capture-avoiding substitution) of this property is the abstraction case, where substitution has to perform $\alpha$-renaming because the
locally abstracted variable happens to be free in the term we are substituting.
However, precisely in that case, we cannot use the induction hypothesis to perform the $\alpha$-renaming because of the precondition on free variables.
Instead, we find it more convenient to prove the substitution lemma using the simple substitution \lstinline|ssubst| and its properties.

For instance, with generalized rewriting in Rocq~\cite{sozeau2009new}, and using the previously defined lemmas, we can replace original terms with their $\vcrefresh$-ed versions, taking
\begin{align*}
  t_1' &:= \vcrefresh\ t_1\ (\FV(u_2) \app \FV(u_1) \app [y; x]), \\
  u_1' &:= \vcrefresh\ u_1\ (\FV(u_2) \app \FV(u_1) \app [y; x]), \\
  u_2' &:= \vcrefresh\ u_2\ (\FV(u_2) \app \FV(u_1) \app [y; x]).
\end{align*}

\noindent
Because the refreshed terms satisfy the required \lstinline|VC| predicates, we can rewrite using \lstinline|vcEq| to replace capture-avoiding substitutions with non-capture-avoiding ones. Thus, it is sufficient to prove
\[\simplesubst\ (\simplesubst\ t_1'\ x\ u_1')\ y\ u_2' \,\eqv[\alpha]\, \simplesubst\ (\simplesubst\ t_1'\ y \ u_2')\ x \ (\simplesubst\ u_1'\ y\ u_2') .\]

However, this can be proved directly by induction on the term $t_1'$ provided that $x\not= y$, $x\notin \textsf{FV}(u_2)$, and $y$ is not a bound variable in $t_1'$. Indeed, variables $x$ and $y$ are different and $x\notin \textsf{FV}(u_2)$ by the assumptions of the substitution lemma. Variable $y$ is not bound in $t_1' = \vcrefresh\ t_1\ (\FV(u_2) \app \FV(u_1) \app [y; x])$ because of the \lstinline|VC| property. Thus, we have proved the substitution lemma for our definition of capture-avoiding substitution \lstinline|subst|.

\section{Comparison with de Bruijn indices\label{s:debruijn}}

De Bruijn indices \cite{de1972lambda} remove the choice of names for bound
variables and thereby turn $\alpha$-equivalence into equality. This observation
leads to two distinct alternatives to the direct $\alpha$-equivalence relation of this paper. One
can use de Bruijn terms as the primary syntax, or one can retain named terms at
the interface and define their $\alpha$-equivalence by conversion to de Bruijn
form. We formalized both alternatives in Rocq in order to compare the proof
obligations.

\subsection{De Bruijn terms as the primary syntax}

When de Bruijn terms are used from the outset, local variables are natural
numbers identifying the abstraction that binds them. Global variables may
retain their names. For example, the syntax can be given by the following \rocqsource{LamDB.v}{62}{datatype}:
\begin{lstlisting}
Inductive Lam : Type :=
  | Var : V -> Lam
  | Rel : nat -> Lam
  | App : Lam -> Lam -> Lam
  | Abs : Lam -> Lam.
\end{lstlisting}
In this representation, $\alpha$-equivalence is simply propositional equality, as in the following \rocqsource{LamDB.v}{79}{definition}:
\begin{lstlisting}
Definition Alpha (t u : Lam) : Prop := t = u.
Infix "=α" := Alpha (at level 60).
\end{lstlisting}
Consequently, reflexivity, symmetry, transitivity, congruence, and preservation
of free variables have immediate proofs. Substitution for global variables and
substitution for a bound index are total functions. Because $\alpha$-equivalence
is equality, the native de Bruijn analogue of \lstinline|SubstAlpha| is
immediate: applying either substitution operation at the same variable or
index to equal source and replacement terms yields equal results by
congruence. The substitution-composition theorem for global variables is an
equation between total functions, proved by a short structural induction;
substitution for relative indices separately requires the usual shifting
operation.

Our native de Bruijn development confirms that this is a concise and attractive
formalization when subsequent definitions and statements also use de Bruijn
terms. It answers a different question, however, from the main development in this paper:
there is then no named input syntax whose substitution and binding judgments
must be justified.

\subsection{Named terms via de Bruijn conversion}

For a comparison on the same named terms as in the rest of the paper, we use a
conversion that maintains a stack of local variable names. On encountering a
variable, \lstinline|findVar| returns its index in that stack, or a named global
variable when it is not present. This development uses a separate de Bruijn
datatype whose constructor names make the distinction from named terms
explicit in the following \rocqsource{LamDBdetour.v}{373}{definition}:
\begin{lstlisting}
Inductive LamDB : Type :=
  | VarDB : V -> LamDB
  | RelDB : nat -> LamDB
  | AppDB : LamDB -> LamDB -> LamDB
  | AbsDB : LamDB -> LamDB.
\end{lstlisting}
The conversion is then given by the following \rocqsource{LamDBdetour.v}{379}{definitions}:
\begin{lstlisting}
Fixpoint findVar (ctx : slist V) (n : nat) (x : V) : LamDB :=
  match ctx with
  | [] => VarDB x
  | ctx' :< y => if x =v? y then RelDB n else findVar ctx' (S n) x
  end.
Equations Lam2DB (ctx : slist V) (t : Lam) : LamDB :=
Lam2DB ctx (Var x) := findVar ctx 0 x;
Lam2DB ctx (App f a) := AppDB (Lam2DB ctx f) (Lam2DB ctx a);
Lam2DB ctx (Abs x b) := AbsDB (Lam2DB (ctx :< x) b).
\end{lstlisting}
The conversion mirrors both our inductive definition and the inductive
relation used by McKinna and Pollack~\cite{mckinna1993pts,mckinna1999some}. It
is also closely related to the decision procedure \lstinline|decVarAlpha| from
Section~\ref{s:adec}.
However, because the context grows in the abstraction case, the translation-induced
relation is again context-indexed, as in the following \rocqsource{LamDBdetour.v}{414}{definition}:
\begin{lstlisting}
Definition Alpha (xs ys : slist V) (t u : Lam) : Prop :=
  Lam2DB xs t = Lam2DB ys u.

Notation "❬ xs ; t ❭ =αd ❬ ys ; u ❭" := (Alpha xs ys t u)
  (at level 70, xs at next level, ys at next level).
Notation "t =αd u" := (Alpha [] [] t u) (at level 70).
\end{lstlisting}
At empty contexts, this gives the usual $\alpha$-equivalence of named terms.
The generalized contextual relation is the one naturally induced by the
translation. In particular, irrelevant outer context entries disappear --- it
need not coincide literally with every generalization of a contextual
inductive judgment when the two context lengths differ.

Equality makes several proofs about this relation immediate. Transitivity and
the ``logical'' axioms of \citet{barendregt1984lambda}, for example, require
little more than equality reasoning. Other results still require a bridge
between named operations and translated terms. For relational substitution,
the central bridge is the following context-aware correspondence lemma, where
\lstinline|substDBCtx| substitutes at a specified de Bruijn index; its statement is the following \rocqsource{LamDBdetour.v}{690}{lemma}:
\begin{lstlisting}
Lemma Subst_Lam2DB_contextual : forall xs xs' x t u v,
  x ∉ xs' -> t⎡x→u⎤~v ->
  substDBCtx (Lam2DB (xs++xs') u) (length xs') (Lam2DB (xs++[x]++xs') t) =
    Lam2DB (xs++xs') v.
\end{lstlisting}
This lemma accounts for shadowing, insertion, and removal of a context entry,
and lifting the substituted term below abstractions. Once it is available, the
de Bruijn version of \lstinline|SubstAlphaMain| is indeed short. Under the same
hypotheses as the theorem in Section~\ref{s:sprop}, its \rocqsource{LamDBdetour.v}{868}{proof} is exactly:
\begin{lstlisting}
move=> t t' u u' v v' xs ys x y xs' ys'
  Hlen Hx Hy Huu Htt Hsub Hsub'.
have Ev := Subst_Lam2DB_contextual xs xs' x t u v Hx Hsub.
have Ev' := Subst_Lam2DB_contextual ys ys' y t' u' v' Hy Hsub'.
unfold Alpha in *. rewrite <- Ev, <- Ev', Huu, Htt, Hlen. reflexivity.
\end{lstlisting}
Thus, conversion does shorten the final compatibility argument. Its premises
are discharged by the correspondence layer: lemmas about lookup under context
insertion, translation under lifting, cancellation of lifting by substitution,
and shadowing.

The executable capture-avoiding substitution theorem still requires much of its
named-term machinery under the detour. Since its inputs are still named, the
development continues to need freshening, the variable convention, and the
connection between relational and executable substitution. In both
formalizations, the functional theorem uses the unfolding equations generated
by Equations and therefore depends on \lstinline|functional_extensionality_dep|.

\subsection{What the comparison shows}

The principal proof obligations are summarized in
Table~\ref{tab:debruijn-comparison}.
\begin{table}[tbp]
\centering
\small
\caption{Comparison of the obligations for the direct named
proofs and the de Bruijn detour.}
\label{tab:debruijn-comparison}
\begin{tabular}{@{}l>{\raggedright\arraybackslash}p{0.36\linewidth}>{\raggedright\arraybackslash}p{0.36\linewidth}@{}}
\toprule
Result & Direct named relation & De Bruijn detour for named terms \\
\midrule
Elementary $\alpha$-laws
  & Short inductions or applications of constructors
  & Mostly equality reasoning \\
\addlinespace
Free variables
  & Context filtering and weakening lemmas
  & Translation preserves named globals \\
\addlinespace
Subst. compatibility
  & Induction on substitution and $\alpha$-derivations
  & A short final equality proof after the substitution-translation lemma \\
\addlinespace
Executable subst.
  & Freshening and the variable convention
  & The same named-term layer, plus its connection to translation \\
\bottomrule
\end{tabular}
\end{table}
The comparison does not establish that either representation is
uniformly shorter or preferable. Native de Bruijn syntax gives the smallest
development when indices are also suitable for the statements of interest.
For named terms, the detour makes the final equality arguments compact, but
moves substantial reasoning into the translation and index algebra. The direct
relation keeps statements, induction, and inversion in the named syntax used
throughout the paper and makes the correspondence between binders explicit.
This is useful for our expository goal, while the formalized detour confirms
that the de Bruijn alternative is also viable.

\section{Conclusion\label{s:conc}}

Our goal in this paper was to investigate whether it is reasonable to formalize $\alpha$-equivalence of untyped $\lambda$-terms in the style of \citet{hindley1986introduction} and \citet{curry1958combinatory}.
To this end, we proposed an inductive definition of $\alpha$-equivalence and showed that it satisfies standard properties expected of $\alpha$-equivalence and other relations used for equational reasoning.
All our definitions and proofs have been formalized in the Rocq Prover.
In addition, our Rocq formalization includes a decision procedure for $\alpha$-equivalence of $\lambda$-terms.
To demonstrate the feasibility of our approach, we also provide a proof of the substitution lemma by explicitly using the variable convention.
Furthermore, we also compare our direct approach to $\alpha$-equivalence with
approaches using de Bruijn indices.


\section*{Acknowledgements}
This work was supported by the European Union and the Estonian Research Council via projects PRG2764 and TEM-TA119.
We thank James McKinna for enlightening discussions and for explaining his and Randy Pollack's work. 
We thank an anonymous reviewer for providing the proof sketch that motivated our expanded comparison with de Bruijn indices.


\renewcommand*{\bibfont}{\normalfont\bibliofont} 
\printbibliography

\pagebreak
\appendix

\section{Rocq Formalization (without proofs)}\label{a:rocq-formalization}

Rocq files and proofs of the following properties (presented as \lstinline|Parameter|s) can be found here: \url{https://github.com/kalmera/lambda-simple-alpha/tree/aug26}. The direct named development is in \lstinline|LamDirect.v|, with list support in \lstinline|MyUtils.v|. The native de Bruijn development, the named-term development via de Bruijn conversion, and the executable examples are in \lstinline|LamDB.v|, \lstinline|LamDBdetour.v|, and \lstinline|LamTest.v|, respectively. The following code has been tested with the Rocq Prover version 9.1.1 together with the Equations plugin version 1.3.1.

\vbadness=10000 

\subsection{Variables} 

The formalizations are parametric over the following module type.

\begin{lstlisting}[language=Coq,breaklines=true,firstline=25,lastline=47,belowskip=0pt]
Module Type LamConf.

(* Parameters *)
Parameter V : Type.
Parameter eqvb : V -> V -> bool.
Parameter eqvb_reflect : ∀ x y, reflect (x = y) (eqvb x y).
Infix "=v?" := eqvb (at level 70).
Parameter fresh : slist V -> V.
Parameter fresh_prop : ∀ xs, fresh xs ∉ xs.

(* Utils *)
Fixpoint in_list x xs : bool :=
  match xs with
  | [] => false
  | ys :< y => if eqvb x y then true else in_list x ys
  end.

Notation "x '∈?' xs" := (in_list x xs)
  (at level 50, no associativity) : type_scope.

Notation "xs -- ys" := (filter (fun y => negb (in_list y ys)) xs)
  (at level 55) : type_scope.
\end{lstlisting}
\begin{lstlisting}[language=Coq,breaklines=true,firstline=56,lastline=56,belowskip=0pt]
End LamConf.
\end{lstlisting}

\subsection{Direct named specification}

This is the specification for direct, inductively defined $\alpha$-equivalence for named lambda terms.

\begin{lstlisting}[language=Coq,breaklines=true,firstline=66,lastline=360,belowskip=0pt]
Module Type LamSpec (Import C : LamConf).

(*
    Basic Definition
*)
Inductive Lam : Type :=
  | Var : V -> Lam
  | App : Lam -> Lam -> Lam
  | Abs : V -> Lam -> Lam.

Equations FV (t : Lam) : slist V :=
FV (Var x) := [x];
FV (App f t) := FV f ++ FV t;
FV (Abs x t) := FV t -- [x].

Equations Vars (t : Lam) : slist V :=
Vars (Var x) := [x];
Vars (App f t) := Vars f ++ Vars t;
Vars (Abs x t) := Vars t :< x.

(*
    Definition of (generic) alpha-equivalence for variables
*)
Reserved Notation "❬ xs ; x ❭ =vα ❬ ys ; y ❭"
    (at level 60, format "'❬' xs ';'  x '❭'  '=vα'  '❬' ys ';'  y '❭'").

Inductive VarAlpha : slist V -> slist V -> V -> V -> Prop :=
  | GlobalVA x : ❬[]; x❭ =vα ❬[]; x❭
  | HereVA xs ys x y : ❬xs :< x; x❭ =vα ❬ys :< y; y❭
  | ThereVA xs ys x y a b :
      x ≠ a -> y ≠ b -> ❬xs; x❭ =vα ❬ys; y❭ -> ❬xs :< a; x❭ =vα ❬ys :< b; y❭
where "❬ xs ; x ❭ =vα ❬ ys ; y ❭" := (VarAlpha xs ys x y).

(*
    Definition of (generic) alpha-equivalence for terms
*)
Reserved Notation "❬ xs ; x ❭ =α ❬ ys ; y ❭"
    (at level 60, format "'❬' xs ';'  x '❭'  '=α'  '❬' ys ';'  y '❭'").

Inductive Alpha : slist V -> slist V -> Lam -> Lam -> Prop :=
  | VarA xs ys x y :
      ❬xs; x❭ =vα ❬ys; y❭ -> ❬xs; Var x❭ =α ❬ys; Var y❭
  | AppA xs ys f g t u :
      ❬xs; f❭ =α ❬ys; g❭ -> ❬xs; t❭ =α ❬ys; u❭ -> ❬xs; App f t❭ =α ❬ys; App g u❭
  | AbsA xs ys x y t u :
      ❬xs :< x; t❭ =α ❬ys :< y; u❭ -> ❬xs; Abs x t❭ =α ❬ys; Abs y u❭
where "❬ xs ; t ❭ =α ❬ ys ; u ❭" := (Alpha xs ys t u).

(* Normal alpha-equivalence *)
Notation "t '=α' u" := (❬[]; t❭ =α ❬[]; u❭) (at level 60).

(* Generic Alpha is an equivalence relation: *)
Parameter AlphaRefl : ∀ t xs, ❬xs; t❭ =α ❬xs; t❭.
Parameter AlphaSym : ∀ t u xs ys, ❬xs; t❭ =α ❬ys; u❭ -> ❬ys; u❭ =α ❬xs; t❭.
Parameter AlphaTrans : ∀ t u v xs ys zs, ❬xs; t❭=α❬ys; u❭ -> ❬ys; u❭=α❬zs; v❭ ->
  ❬xs; t❭=α❬zs; v❭. 


(* Free variables of alpha-equivalent terms are same *)
Parameter AlphaFV : ∀ t u xs ys, ❬xs; t❭ =α ❬ys; u❭ -> FV t -- xs = FV u -- ys.

(* Barendregt&Barendsen style compatibility rules:
*)
Parameter AlphaConvAppFun : ∀ (t u v : Lam), t =α u -> App t v =α App u v.
Parameter AlphaConvAppArg : ∀ (t u v : Lam), t =α u -> App v t =α App v u.
Parameter AlphaConvAbs : ∀ (t u : Lam) (x : V), t =α u -> Abs x t =α Abs x u.

(* Removing of shadowed variables *)
Parameter AlphaRemShadowed :
  ∀ t xs3 ys3 u x x' y y' xs2 xs1 ys2 ys1,
  (x = x' \/ y = y') ->
  length xs3 = length ys3 ->
  length xs2 = length ys2 ->
  ❬xs1++[x']++xs2++[x]++xs3; t❭ =α ❬ys1++[y']++ys2++[y]++ys3; u❭ ->
  ❬xs1++xs2++[x]++xs3; t❭ =α ❬ys1++ys2++[y]++ys3; u❭.

(* Making implicit globals explicit or explicit globals implicit *)
Parameter AlphaCtxImplGlobs : ∀ t u xs ys zs,
  length xs = length ys ->
  ❬xs; t❭ =α ❬ys; u❭ <-> ❬zs++xs; t❭ =α ❬zs++ys; u❭.

(* Context weakening rule (unrelated variables can be added) *)
Parameter AlphaCtxWeaken :
  ∀ t u xs xs' ys ys' x y,
  length xs = length ys ->
  ❬xs' ++ xs; t❭ =α ❬ys' ++ ys; u❭ ->
  (x ∉ FV t \/ x ∈ xs) /\ (y ∉ FV u \/ y ∈ ys) ->
  ❬xs' ++ [x] ++ xs; t❭ =α ❬ys' ++ [y] ++ ys; u❭.

(* variable exchange rule *)
Parameter AlphaSwap :
  ∀ t u xs ys xs' ys' a x b y,
  length xs = length ys ->
  ❬xs'++[a; x]++xs; t❭ =α ❬ys'++[b; y]++ys; u❭ ->
  a ≠ x -> b ≠ y ->
  ❬xs'++[x; a]++xs; t❭ =α ❬ys'++[y; b]++ys; u❭.


(*
    Definition and properties of substitutions
*)

Reserved Notation "t '⎡' y '→' u '⎤' '~' v"
    (at level 60, format "t '⎡' y '→' u '⎤'  '~'  v").

(* capture-avoiding substitution -- as a relation *)
Inductive Subst (x : V) (u : Lam) : Lam -> Lam -> Prop :=
  | SubstVarEq : (Var x)⎡x→u⎤~u
  | SubstVarNeq y : y ≠ x -> (Var y)⎡x→u⎤~Var y
  | SubstApp f f' t t' : f⎡x→u⎤~f' -> t⎡x→u⎤~t' -> (App f t)⎡x→u⎤~(App f' t')
  | SubstAbsEq t : (Abs x t)⎡x→u⎤~(Abs x t)
  | SubstAbsNF y t t' :
      y ≠ x -> y ∉ FV u -> t⎡x→u⎤~t' -> (Abs y t)⎡x→u⎤~(Abs y t')
where "t ⎡ y → u ⎤ ~ v" := (Subst y u t v).


(* non-capture-avoiding substitution -- as a function *)
Equations ssubst (t : Lam) (x : V) (u : Lam) : Lam :=
ssubst (Var y) x u := if y =v? x then u else Var y;
ssubst (App f t) x u := App (ssubst f x u) (ssubst t x u);
ssubst (Abs y t) x u :=
  if y =v? x then Abs y t else Abs y (ssubst t x u).


(* Barendregt&Barendsen style alpha-rule: *)
Parameter SubstFreshVarAlpha : ∀ t t' x y, 
    y ∉ FV t -> t⎡x→Var y⎤~t' -> Abs x t =α Abs y t'.

(* Substitution does not change t if variable does not occur free in t *)
Parameter AlphaSubstUnusedVar :
  ∀ t x u v, x ∉ FV t -> t⎡x→u⎤~v -> t = v.

(* General alpha-equivalence of substitution. *)
Parameter SubstAlpha :
  ∀ t t' u u' v v' xs ys x y,
  ❬xs; u❭ =α ❬ys; u'❭ ->
  ❬xs; Abs x t❭ =α ❬ys; Abs y t'❭ ->
  t⎡x→u⎤~v ->
  t'⎡y→u'⎤~v' ->
  ❬xs; v❭ =α ❬ys; v'❭.

(* One of the main (and most difficult) substitution properties.  *)
Parameter SubstAlphaMain :
  ∀ t t' u u' v v' xs ys x y xs' ys',
  length xs' = length ys' ->
  x ∉ xs' -> y ∉ ys' ->
  ❬xs++xs'; u❭ =α ❬ys++ys'; u'❭ ->
  ❬xs++[x]++xs'; t❭ =α ❬ys++[y]++ys'; t'❭ ->
  t⎡x→u⎤~v ->
  t'⎡y→u'⎤~v' ->
  ❬xs++xs'; v❭ =α ❬ys++ys'; v'❭.

(* Corollary of SubstAlpha *)
Parameter Contraction :
  ∀ t t' v v' xs ys xs' ys' x1 x2 y1 y2,
  length xs = length ys ->
  x1 ∉ xs -> y1 ∉ ys ->
  x2 ∉ ([x1] ++ xs) -> y2 ∉ ([y1] ++ ys) ->
  ❬xs' ++ [x2; x1] ++ xs; t❭ =α ❬ys' ++ [y2; y1] ++ ys; t'❭ ->
  t⎡x2→Var x1⎤~v ->
  t'⎡y2→Var y1⎤~v' ->
  ❬xs' ++ [x1] ++ xs; v❭ =α ❬ys' ++ [y1] ++ ys; v'❭.

(* Corollary of SubstAlpha *)
Parameter Substitutivity :
  ∀ t x u u' v v',
  u =α u' -> t⎡x→u⎤~v -> t⎡x→u'⎤~v' -> v =α v'.

(* Corollary of SubstAlphaMain *)
Parameter Congruence :
  ∀ t t' u v v' xs ys x y xs' ys',
  length xs' = length ys' ->
  x ∉ xs' -> y ∉ ys' ->
  ❬xs ++ xs'; u❭ =α ❬ys ++ ys'; u❭ ->
  ❬xs ++ [x] ++ xs'; t❭ =α ❬ys ++ [y] ++ ys'; t'❭ ->
  t⎡x→u⎤~v -> t'⎡y→u⎤~v' ->
  ❬xs ++ xs'; v❭ =α ❬ys ++ ys'; v'❭.


(*
    Implementation of capture-avoiding substitutions
*)

Equations height (t : Lam) : nat :=
height (Var _) := 1;
height (App f t) := 1 + max (height f) (height t);
height (Abs _ t) := 1 + height t.


Parameter ssubst_height : ∀ t x z, height (ssubst t x (Var z)) = height t.

Equations? subst (t : Lam) (x : V) (u : Lam) : Lam by wf (height t) :=
subst (Var v) x u :=
  if v =v? x then u else Var v;

subst (App f t) x u :=
  App (subst f x u) (subst t x u);

subst (Abs y t) x u :=
  if y =v? x then Abs y t
  else if y ∈? FV u then
    let z  := fresh (Vars t ++ FV u ++ [y; x]) in
    let t' := ssubst t y (Var z) in
    Abs z (subst t' x u)
  else Abs y (subst t x u).
Proof.
  all: simp height; try lia.
  unfold t'; clear t'.
  rewrite ssubst_height. lia.
Qed.

(* subst conforms to our specification *)
Parameter SubstExists : ∀ t u x,
  { t' : Lam | t' =α t & t'⎡x→u⎤~subst t x u }.


(*
    Decision proc. of alpha-equivalence
*)

Equations decVarAlpha (xs ys : slist V) (x y : V) : bool :=
decVarAlpha (xs' :< x') (ys' :< y') x y :=
  match x =v? x', y =v? y' with
  | true, true => true
  | false, false => decVarAlpha xs' ys' x y
  | _, _ => false
  end;

decVarAlpha [] [] x y := x =v? y;
decVarAlpha _ _ _ _ := false.

Equations decAlpha (xs ys : slist V) (t u : Lam) : bool :=
decAlpha xs ys (Var x) (Var y) := decVarAlpha xs ys x y;
decAlpha xs ys (App f t) (App g u) := decAlpha xs ys f g && decAlpha xs ys t u;
decAlpha xs ys (Abs x t) (Abs y u) := decAlpha (xs :< x) (ys :< y) t u;
decAlpha xs ys _ _ => false.

(* Specification: decAlpha returns true iff alpha-equivalent *)
Parameter decAlphaSpec : ∀ (xs ys : slist V) (t u : Lam),
  decAlpha xs ys t u = true <-> ❬xs; t❭ =α ❬ys; u❭.

(*
    (Barendregt) Variable Condition
*)
Inductive VC (xs : slist V) : Lam -> Prop :=
  | VCVar x : VC xs (Var x)
  | VCApp f t : VC xs f -> VC xs t -> VC xs (App f t)
  | VCAbs x t : x ∉ xs -> VC xs t -> VC xs (Abs x t).

(* Condtition on when simple substitution suffices *)
Parameter vcEq :
  ∀ t x xs v, FV v ⊆ xs -> VC xs t -> subst t x v =α ssubst t x v.

(*
    Variable Condition refresh

    "vcRefresh t vs" returns a term t' such that
        * t' =α t
        * VC vs t'
*)
Equations? vcRefresh (t : Lam) (xs : slist V) : Lam by wf (height t) :=
vcRefresh (Var x) _ :=
    Var x;

vcRefresh (App f u) xs :=
    App (vcRefresh f xs) (vcRefresh u xs);

vcRefresh (Abs y u) xs :=
    if y ∈? xs then
        let y' := fresh (Vars u ++ xs :< y) in
        let u' := ssubst u y (Var y') in
        Abs y' (vcRefresh u' xs)
    else
        Abs y (vcRefresh u xs).
Proof.
    all: simp height; try lia.
    rewrite ssubst_height. lia.
Qed.

(* vcRefresh returns an alpha-equivalent term *)
Parameter AlphaVcRefresh : ∀ t xs, t =α vcRefresh t xs.

(* vcRefresh t vs satisfies the variable convention *)
Parameter vcRefreshProp : ∀ t xs, VC xs (vcRefresh t xs).

(* ssubst does not require a refresh *)
Parameter VCssubst : ∀ t x xs u,
  VC xs u -> VC xs t -> VC xs (ssubst t x u).

(* The substitution lemma. *)
Parameter SubstSwap :
  ∀ t x u y v,
  x ≠ y -> x ∉ FV v ->
  subst (subst t x u) y v =α subst (subst t y v) x (subst u y v).
End LamSpec.
\end{lstlisting}

\subsection{Named terms via the de Bruijn detour}

This is the alternative specification for named lambda terms where $\alpha$-equivalence is defined by equality of their de Bruijn index forms.

\begin{lstlisting}[language=Coq,breaklines=true,firstline=77,lastline=328,belowskip=0pt]
Module Type LamDBDetourSpec (Import C : LamConf).

(** * Named syntax and relational substitution *)

Inductive Lam : Type :=
  | Var : V -> Lam
  | App : Lam -> Lam -> Lam
  | Abs : V -> Lam -> Lam.

Equations FV (t : Lam) : slist V :=
FV (Var x) := [x];
FV (App f a) := FV f ++ FV a;
FV (Abs x b) := FV b -- [x].

Equations Vars (t : Lam) : slist V :=
Vars (Var x) := [x];
Vars (App f a) := Vars f ++ Vars a;
Vars (Abs x b) := Vars b :< x.

Reserved Notation "t '⎡' x '→' u '⎤' '~' v"
  (at level 60, format "t '⎡' x '→' u '⎤'  '~'  v").

Inductive Subst (x : V) (u : Lam) : Lam -> Lam -> Prop :=
  | SubstVarEq : Var x⎡x→u⎤~u
  | SubstVarNeq y : y ≠ x -> Var y⎡x→u⎤~Var y
  | SubstApp f f' a a' :
      f⎡x→u⎤~f' -> a⎡x→u⎤~a' ->
      App f a⎡x→u⎤~App f' a'
  | SubstAbsEq b : Abs x b⎡x→u⎤~Abs x b
  | SubstAbsFresh y b b' :
      y ≠ x -> y ∉ FV u -> b⎡x→u⎤~b' ->
      Abs y b⎡x→u⎤~Abs y b'
where "t ⎡ x → u ⎤ ~ v" := (Subst x u t v).


(** * De Bruijn translation and alpha-equivalence *)

Inductive LamDB : Type :=
  | VarDB : V -> LamDB
  | RelDB : nat -> LamDB
  | AppDB : LamDB -> LamDB -> LamDB
  | AbsDB : LamDB -> LamDB.

Fixpoint findVar (ctx : slist V) (n : nat) (x : V) : LamDB :=
  match ctx with
  | [] => VarDB x
  | ctx' :< y => if x =v? y then RelDB n else findVar ctx' (S n) x
  end.

Equations Lam2DB (ctx : slist V) (t : Lam) : LamDB :=
Lam2DB ctx (Var x) := findVar ctx 0 x;
Lam2DB ctx (App f a) := AppDB (Lam2DB ctx f) (Lam2DB ctx a);
Lam2DB ctx (Abs x b) := AbsDB (Lam2DB (ctx :< x) b).

Fixpoint lift (d k : nat) (t : LamDB) : LamDB :=
  match t with
  | VarDB x => VarDB x
  | RelDB i => if i <? k then RelDB i else RelDB (d + i)
  | AppDB f a => AppDB (lift d k f) (lift d k a)
  | AbsDB b => AbsDB (lift d (S k) b)
  end.

Fixpoint substDBCtx (u : LamDB) (k : nat) (t : LamDB) : LamDB :=
  match t with
  | VarDB x => VarDB x
  | RelDB i =>
      if i <? k then RelDB i
      else if i =? k then u else RelDB (i - 1)
  | AppDB f a => AppDB (substDBCtx u k f) (substDBCtx u k a)
  | AbsDB b => AbsDB (substDBCtx (lift 1 0 u) (S k) b)
  end.

Definition Alpha (xs ys : slist V) (t u : Lam) : Prop :=
  Lam2DB xs t = Lam2DB ys u.

Notation "❬ xs ; t ❭ =αd ❬ ys ; u ❭" := (Alpha xs ys t u)
  (at level 70, xs at next level, ys at next level).
Notation "t =αd u" := (Alpha [] [] t u) (at level 70).

Parameter AlphaRefl : forall xs t, ❬xs; t❭ =αd ❬xs; t❭.
Parameter AlphaSym : forall xs ys t u,
  ❬xs; t❭ =αd ❬ys; u❭ -> ❬ys; u❭ =αd ❬xs; t❭.
Parameter AlphaTrans : forall xs ys zs t u v,
  ❬xs; t❭ =αd ❬ys; u❭ -> ❬ys; u❭ =αd ❬zs; v❭ ->
  ❬xs; t❭ =αd ❬zs; v❭.
Parameter AlphaFV : forall xs ys t u,
  ❬xs; t❭ =αd ❬ys; u❭ -> FV t -- xs = FV u -- ys.
Parameter AlphaConvApp : forall xs ys f g a b,
  ❬xs; f❭ =αd ❬ys; g❭ -> ❬xs; a❭ =αd ❬ys; b❭ ->
  ❬xs; App f a❭ =αd ❬ys; App g b❭.
Parameter AlphaConvAbsCtx : forall xs ys x y t u,
  ❬xs :< x; t❭ =αd ❬ys :< y; u❭ ->
  ❬xs; Abs x t❭ =αd ❬ys; Abs y u❭.
Parameter AlphaConvAbs : forall x t u,
  t =αd u -> Abs x t =αd Abs x u.


(** * Substitution compatibility *)

Parameter Subst_Lam2DB_contextual : forall xs xs' x t u v,
  x ∉ xs' ->
  t⎡x→u⎤~v ->
  substDBCtx (Lam2DB (xs ++ xs') u) (length xs') (Lam2DB (xs ++ [x] ++ xs') t) = 
    Lam2DB (xs ++ xs') v.

Parameter SubstAlphaMain :
  forall t t' u u' v v' xs ys x y xs' ys',
  length xs' = length ys' ->
  x ∉ xs' -> y ∉ ys' ->
  ❬xs ++ xs'; u❭ =αd ❬ys ++ ys'; u'❭ ->
  ❬xs ++ [x] ++ xs'; t❭ =αd ❬ys ++ [y] ++ ys'; t'❭ ->
  t⎡x→u⎤~v -> t'⎡y→u'⎤~v' ->
  ❬xs ++ xs'; v❭ =αd ❬ys ++ ys'; v'❭.

Parameter SubstAlpha : forall t t' u u' v v' xs ys x y,
  ❬xs; u❭ =αd ❬ys; u'❭ ->
  ❬xs; Abs x t❭ =αd ❬ys; Abs y t'❭ ->
  t⎡x→u⎤~v -> t'⎡y→u'⎤~v' ->
  ❬xs; v❭ =αd ❬ys; v'❭.

Parameter SubstAlpha_Simple : forall t t' x u v v',
  t =αd t' -> t⎡x→u⎤~v -> t'⎡x→u⎤~v' -> v =αd v'.

Parameter Substitutivity : forall t x u u' v v',
  u =αd u' -> t⎡x→u⎤~v -> t⎡x→u'⎤~v' -> v =αd v'.

Parameter Congruence : forall t t' u v v' xs ys x y xs' ys',
  length xs' = length ys' ->
  x ∉ xs' -> y ∉ ys' ->
  ❬xs ++ xs'; u❭ =αd ❬ys ++ ys'; u❭ ->
  ❬xs ++ [x] ++ xs'; t❭ =αd ❬ys ++ [y] ++ ys'; t'❭ ->
  t⎡x→u⎤~v -> t'⎡y→u⎤~v' ->
  ❬xs ++ xs'; v❭ =αd ❬ys ++ ys'; v'❭.

Parameter Contraction :
  forall t t' v v' xs ys xs' ys' x1 x2 y1 y2,
  length xs = length ys ->
  x1 ∉ xs -> y1 ∉ ys ->
  x2 ∉ ([x1] ++ xs) -> y2 ∉ ([y1] ++ ys) ->
  ❬xs' ++ [x2; x1] ++ xs; t❭ =αd
    ❬ys' ++ [y2; y1] ++ ys; t'❭ ->
  t⎡x2→Var x1⎤~v -> t'⎡y2→Var y1⎤~v' ->
  ❬xs' ++ [x1] ++ xs; v❭ =αd ❬ys' ++ [y1] ++ ys; v'❭.

Parameter SubstFunctional : forall t x u v v',
  t⎡x→u⎤~v -> t⎡x→u⎤~v' -> v =αd v'.

Parameter substDBCtx_swap : forall t u v k,
  substDBCtx v k (substDBCtx u (S k) t) =
  substDBCtx (substDBCtx v k u) k
    (substDBCtx (lift 1 k v) k t).

Parameter SubstSwapRel :
  forall t1 t2 t3 u1 u2 x y u1' t2' t3',
  x ≠ y -> x ∉ FV u2 ->
  t1⎡x→u1⎤~t2 -> t2⎡y→u2⎤~t3 ->
  u1⎡y→u2⎤~u1' ->
  t1⎡y→u2⎤~t2' -> t2'⎡x→u1'⎤~t3' ->
  t3 =αd t3'.


(** * Decision procedure *)

Fixpoint eqDB (t u : LamDB) : bool :=
  match t, u with
  | VarDB x, VarDB y => x =v? y
  | RelDB i, RelDB j => Nat.eqb i j
  | AppDB f a, AppDB g b => eqDB f g && eqDB a b
  | AbsDB b, AbsDB c => eqDB b c
  | _, _ => false
  end.

Definition decAlpha xs ys t u := eqDB (Lam2DB xs t) (Lam2DB ys u).

Parameter decAlphaSpec : forall xs ys t u,
  decAlpha xs ys t u = true <-> ❬xs; t❭ =αd ❬ys; u❭.


(** * Executable substitution and the variable convention *)

Equations height (t : Lam) : nat :=
height (Var _) := 1;
height (App f a) := 1 + max (height f) (height a);
height (Abs _ b) := 1 + height b.

Equations ssubst (t : Lam) (x : V) (u : Lam) : Lam :=
ssubst (Var y) x u := if y =v? x then u else Var y;
ssubst (App f a) x u := App (ssubst f x u) (ssubst a x u);
ssubst (Abs y b) x u :=
  if y =v? x then Abs y b else Abs y (ssubst b x u).

Parameter ssubst_height : forall t x z,
  height (ssubst t x (Var z)) = height t.

Equations? subst (t : Lam) (x : V) (u : Lam) : Lam by wf (height t) :=
subst (Var y) x u := if y =v? x then u else Var y;
subst (App f a) x u := App (subst f x u) (subst a x u);
subst (Abs y b) x u :=
  if y =v? x then Abs y b
  else if y ∈? FV u then
    let z := fresh (Vars b ++ FV u ++ [y; x]) in
    Abs z (subst (ssubst b y (Var z)) x u)
  else Abs y (subst b x u).
Proof.
  all: simp height; try lia.
  rewrite ssubst_height. lia.
Qed.

Inductive VC (xs : slist V) : Lam -> Prop :=
  | VCVar x : VC xs (Var x)
  | VCApp f a : VC xs f -> VC xs a -> VC xs (App f a)
  | VCAbs x b : x ∉ xs -> VC xs b -> VC xs (Abs x b).

Parameter VCssubst : forall t x xs u,
  VC xs u -> VC xs t -> VC xs (ssubst t x u).

Equations? vcRefresh (t : Lam) (avoid : slist V) : Lam by wf (height t) :=
vcRefresh (Var x) _ := Var x;
vcRefresh (App f a) avoid :=
  App (vcRefresh f avoid) (vcRefresh a avoid);
vcRefresh (Abs x b) avoid :=
  if x ∈? avoid then
    let y := fresh ((Vars b ++ avoid) :< x) in
    Abs y (vcRefresh (ssubst b x (Var y)) avoid)
  else Abs x (vcRefresh b avoid).
Proof.
  all: simp height; try lia.
  rewrite ssubst_height. lia.
Qed.

Parameter AlphaVcRefresh : forall t avoid, t =αd vcRefresh t avoid.
Parameter vcRefreshProp : forall t avoid, VC avoid (vcRefresh t avoid).
Parameter vcEq : forall t x xs u,
  FV u ⊆ xs -> VC xs t -> subst t x u =αd ssubst t x u.

Parameter SubstFreshVarAlpha : forall t t' x y,
  y ∉ FV t -> t⎡x→Var y⎤~t' -> Abs x t =αd Abs y t'.
Parameter AlphaSubstUnusedVar : forall t x u v,
  x ∉ FV t -> t⎡x→u⎤~v -> t = v.
Parameter AlphaSubstFun : forall t t' x u u',
  u =αd u' -> t =αd t' -> subst t x u =αd subst t' x u'.
Parameter Subst2SubstFn : forall t x u v,
  t⎡x→u⎤~v -> subst t x u =αd v.
Parameter SubstExists : forall t u x,
  { t' : Lam | t' =αd t & t'⎡x→u⎤~subst t x u }.

Parameter SubstSwap : forall t x u y v,
  x ≠ y -> x ∉ FV v ->
  subst (subst t x u) y v =αd subst (subst t y v) x (subst u y v).

End LamDBDetourSpec.
\end{lstlisting}
\end{document}